\documentclass{revtex4-2}%
\usepackage[latin9]{inputenc}
\usepackage{amsmath}
\usepackage{amssymb}
\usepackage{graphicx}
\usepackage[unicode=true,  bookmarks=false,  breaklinks=false,pdfborder={0 0
1},backref=section,colorlinks=false]{hyperref}
\usepackage{amsfonts}
\usepackage{subfigure}
\usepackage{cleveref}
\usepackage{listings}
\usepackage{xcolor}
\usepackage{array}
\usepackage{url}%
\hypersetup{
colorlinks,linkcolor=blue,citecolor=blue,urlcolor=blue}
\makeatletter
\@ifundefined{textcolor}{}
{
\definecolor{BLACK}{gray}{0}
\definecolor{WHITE}{gray}{1}
\definecolor{RED}{rgb}{1,0,0}
\definecolor{GREEN}{rgb}{0,1,0}
\definecolor{BLUE}{rgb}{0,0,1}
\definecolor{CYAN}{cmyk}{1,0,0,0}
\definecolor{MAGENTA}{cmyk}{0,1,0,0}
\definecolor{YELLOW}{cmyk}{0,0,1,0}
}

\makeatother
\begin{document}
\preprint{CTP-SCU/2021018}
\title{Photon Ring and Observational Appearance of a Hairy Black Hole}
\author{Qingyu Gan}
\email{gqy@stu.scu.edu.cn}
\author{Peng Wang}
\email{pengw@scu.edu.cn}
\author{Houwen Wu}
\email{iverwu@scu.edu.cn}
\author{Haitang Yang}
\email{hyanga@scu.edu.cn}
\affiliation{Center for Theoretical Physics, College of Physics, Sichuan University,
Chengdu, 610064, China}

\begin{abstract}
Recently, the image of a Schwarzschild black hole with an accretion disk has
been revisited, and it showed that the \textquotedblleft photon
ring\textquotedblright, defined as highly bent light rays that intersect the
disk plane more than twice, is extremely narrow and makes a negligible
contribution to the total brightness. In this paper, we investigate the
observational appearance of an optically and geometrically thin accretion disk
around a hairy black hole in an Einstein-Maxwell-scalar model. Intriguingly,
we find that in a certain parameter regime, due to an extra maximum or an
\textquotedblleft ankle-like\textquotedblright\ structure in the effective
potential for photons, the photon ring can be remarkably wide, thus making a
notable contribution to the flux of the observed image. In particular, there
appears a wide and bright annulus, which comprises multiple concentric bright
thin rings with different luminosity, in the high resolution image.

\end{abstract}
\maketitle
\tableofcontents
\preprint{CTP-SCU/2021XXX}
\affiliation{Center for Theoretical Physics, College of Physics, Sichuan University,
Chengdu, 610064, PR China}

{}

\section{Introduction}

The announcement of the first angular resolution image of the supermassive
black hole M87{*} by the Event Horizon Telescope (EHT) collaboration is a
significant event in observing astrophysical black holes
\cite{Akiyama:2019cqa,Akiyama:2019brx,Akiyama:2019sww,Akiyama:2019bqs,Akiyama:2019fyp,Akiyama:2019eap,Akiyama:2021qum,Akiyama:2021tfw}%
, which opens a new window to test general relativity in the strong field
regime. The image contains two prominent features, a circular dark interior,
dubbed \textquotedblleft shadow\textquotedblright, and a bright ring which is
closely relevant to a class of circular photon orbits (i.e., photon sphere).
The shadow and photon sphere are originated from the light deflection by the
strong gravitational field near the black hole
\cite{Synge:1966okc,Bardeen:1972fi,Bardeen:1973tla,Bozza:2009yw}. Thus, it is
believed that the shadow image encodes valuable information of the geometry
around the black hole, especially in the vicinity of the horizon.

Modelling the M87{*} with the Kerr black hole geometry, the observation was
found to be in good agreement with the prediction of general relativity.
Nevertheless, due to finite resolution of the M87{*} image, there still exists
some space for alternative models to simulate the black hole image within
observational uncertainty tolerance. To explore these possibilities, one can
parameterize the deviations from the Kerr metric and compare the corresponding
shadow image with the observed image
\cite{Li:2013jra,Tsukamoto:2014tja,Abdujabbarov:2016hnw,Tsukamoto:2017fxq,Kumar:2018ple,Dymnikova:2019vuz}%
. Alternatively, shadows and photon spheres of black holes are widely studied
in the context of various specific theories including new physics, e.g., the
nonlinear electrodynamics
\cite{Atamurotov:2015xfa,Stuchlik:2019uvf,Ma:2020dhv,Hu:2020usx,Kruglov:2020tes}%
, the Gauss-Bonnet theory
\cite{Ma:2019ybz,Wei:2020ght,Zeng:2020dco,Guo:2020zmf}, fuzzball
\cite{Bacchini:2021fig}, the Chern-Simons type theory
\cite{Ayzenberg:2018jip,Amarilla:2010zq}, $f(R)$ gravity
\cite{Dastan:2016vhb,Addazi:2021pty}, and string inspired black holes
\cite{Amarilla:2011fx,Guo:2019lur,Kumar:2020hgm,Zhu:2019ura}. Moreover, there
could exist some exotic ultra-compact objects acting as black hole mimickers
\cite{Cunha:2018acu}. The gravitational lensing by various horizonless
objects, such as wormholes \cite{Shaikh:2018oul,Wielgus:2020uqz,Peng:2021osd}
and bosonic stars \cite{Cunha:2017wao,Olivares:2018abq}, has been detailedly
analyzed. Their shadow images are usually distinct from those of black holes,
but it is hard to distinguish between them at the current EHT resolution.
Interestingly, it was argued in
\cite{Shaikh:2018lcc,Joshi:2020tlq,Dey:2020bgo} that naked singularities can
cast a shadow in the absence of the photon sphere. Furthermore, the EHT
observation can also be applied to impose constraints on the cosmological
parameters
\cite{Perlick:2018iye,Tsupko:2019pzg,Vagnozzi:2020quf,Li:2020drn,Roy:2020dyy,Chowdhuri:2020ipb}
and the size of extra dimensions
\cite{Amir:2017slq,Vagnozzi:2019apd,Banerjee:2019nnj}, test the equivalence
principle \cite{Li:2019lsm,Gralla:2020pra,Li:2021mzq}, and probe some
fundamental physics issues including dark matter
\cite{Chen:2019fsq,Konoplya:2019sns,Jusufi2019,Davoudiasl2019,Roy:2019esk,Ma:2020dhv,Saurabh:2020zqg}
and dark energy
\cite{Abdujabbarov:2015pqp,Zeng:2020vsj,Qin:2020xzu,Yuan-Chen:2020sek}.

On the other hand, an astrophysical black hole is generally believed to be
surrounded by a luminous accretion flow, which is an essential ingredient in
obtaining the black hole image. In fact, the realistic image is a result of
the complex interactions between the strong gravitational lensing of the black
hole and the electromagnetic plasma in the accretion flow, which requires
intensive numerical general relativistic magnetohydrodynamic (GRMHD)
simulations \cite{Porth:2019wxk}. Nevertheless, simplified accretion models
usually suffice to capture major features of black hole images, and hence have
been widely investigated in the literature, e.g., spherical accretion flows
\cite{Shaikh:2018lcc,Zeng:2020dco,Zeng:2020vsj,Qin:2020xzu,Narayan:2019imo,Saurabh:2020zqg}%
, thin or thick accretion disks
\cite{Luminet:1979nyg,Beckwith:2004ae,Gralla:2019xty,Dokuchaev:2019pcx,Peng:2020wun,He:2021htq,Eichhorn:2021iwq,Li:2021riw}%
. In particular, the authors in \cite{Gralla:2019xty} considered the emission
from an optically and geometrically thin disk around a Schwarzschild black
hole, which is divided into three classes by the number $m$ of half-orbits
that an emitted photon completes around the black hole before reaching the
observer: the direct emission ($m\leq1$), the lensing ring ($m=2$) and the
photon ring ($m\geq3$). It showed that the lensing ring superimposed upon the
direct emission produces a thin ring of twice the background intensity in the
black hole image, while the photon ring, which picks up larger intensity,
makes negligible contributions to the total observed brightness due to its
exponential narrowness. Nevertheless, observations of the photon ring would
provide a new and powerful tool to probe a black hole spacetime
\cite{Gralla:2019drh}. Recently, experimental methods have been proposed to
detect the photon ring by measuring its interferometric signatures
\cite{Johnson:2019ljv} and two-point correlation function of intensity
fluctuations \cite{Hadar:2020fda}.

No-hair theorem states that a black hole can be completely characterized by
only three observable classical parameters: mass, electric charge, and spin
\cite{Ruffini:1971bza}. However, various models of hairy black holes (HBHs)
have been proposed to circumvent the no-hair theorem (see
\cite{Herdeiro:2015waa} for a review). Testing the no-hair theorem with
observations would provide powerful probes of alternative theories of gravity.
The observation of the black hole shadow would enable tests of the no-hair
theorem, and therefore studying shadows of HBHs with various hairs has
attracted great attention, e.g., axion-like hairs
\cite{Wei:2013kza,Banerjee:2019xds}, dilaton-like hairs
\cite{Amarilla:2013sj,Dastan:2016bfy,Mizuno:2018lxz} and others
\cite{Cunha:2015yba,Cunha:2016bjh,Cunha:2016bpi,Konoplya:2019goy,Konoplya:2019fpy,Khodadi:2020jij,Gao:2021luq}%
. Interestingly, in \cite{Gan:2021pwu} we found that there can exist two
unstable photon spheres outside the event horizon for the HBH solutions in the
Einstein-Maxwell-scalar (EMS) theory proposed in \cite{Herdeiro:2018wub}. This
novel feature results in two concentric bright rings of different radii in the
observed image of the HBH surrounded by an optically thin, spherical accretion
flow. Note that the presence of more than one photon sphere has also been
reported in other scenarios, e.g., horizonless ultra-compact objects
\cite{Cunha:2017qtt}, Morris-Thorne type wormholes
\cite{Shaikh:2018oul,Shaikh:2019jfr}, reflection-asymmetric wormholes
\cite{Wielgus:2020uqz,Peng:2021osd,Tsukamoto:2021fpp,Guerrero:2021pxt}, and a
Schwarzschild black hole surrounded by a certain matter distribution
\cite{Reji:2019brv}.

To investigate the effects of two unstable photon spheres on the photon ring,
we consider the observational appearance of the aforementioned HBH surrounded
by an optically and geometrically thin accretion disk. Remarkably, we show
that, in a certain parameter regime, the existence of two photon spheres or
its reminiscence can significantly extend the photon ring band, which thus
makes a non-trivial contribution to the total observed intensity, comparable
to that of the lensing ring. The rest of the paper is organized as follows. In
Sec. \ref{sec:EMS BH}, we briefly review the HBH solutions, the behavior of
null geodesics and the observed intensity of the accretion disk. Sec.
\ref{sec:numerical result} contains our main numerical results, which include
effective potentials for photons, accretion disk images seen by a distant
observer and the dependence of the photon ring on the HBH charge and scalar
coupling. We conclude with a brief discussion in Sec.
\ref{sec:Discussion-and-conclusion}. We set $16\pi G=1$ throughout the paper.

\section{Set Up}

\label{sec:EMS BH}

Consider a specific 4D EMS theory with an exponential scalar-electromagnetic
coupling given by \cite{Herdeiro:2018wub}
\begin{equation}
S=\int d^{4}x\sqrt{-g}\left[  \mathcal{R}-2\partial_{\mu}\phi\partial^{\mu
}\phi-e^{\alpha\phi^{2}}F_{\mu\nu}F^{\mu\nu}\right]  , \label{eq:action}%
\end{equation}
where $\mathcal{R}$ is the Ricci scalar, the scalar field $\phi$ is minimally
coupled to the metric $g_{\mu\nu}$ and non-minimally coupled to the
electromagnetic field $A_{\mu}$, and $F_{\mu\nu}=\partial_{\mu}A_{\nu
}-\partial_{\nu}A_{\mu}$ is the electromagnetic tensor. Many properties of
this model and its extensions have been explored in the literature, e.g.,
various non-minimal coupling functions
\cite{Fernandes:2019rez,Blazquez-Salcedo:2020nhs}, dyons including magnetic
charges \cite{Astefanesei:2019pfq}, axionic-type couplings
\cite{Fernandes:2019kmh}, massive and self-interacting scalar fields
\cite{Zou:2019bpt,Fernandes:2020gay}, horizonless reflecting stars
\cite{Peng:2019cmm}, stability analysis of the HBHs
\cite{Myung:2018vug,Myung:2019oua,Zou:2020zxq,Myung:2020etf,Mai:2020sac},
higher dimensional scalar-tensor models \cite{Astefanesei:2020qxk},
quasinormal modes of the HBHs \cite{Myung:2018jvi,Blazquez-Salcedo:2020jee},
two U(1) fields \cite{Myung:2020dqt}, quasi-topological electromagnetism
\cite{Myung:2020ctt}, topology and spacetime structure influences
\cite{Guo:2020zqm}, the Einstein-Born-Infeld-scalar theory \cite{Wang:2020ohb}
and with a negative cosmological constant \cite{Guo:2021zed,Zhang:2021etr}.
Starting with the static and spherically symmetric black hole solution
ansatz,
\begin{equation}
ds^{2}=-N(r)e^{-2\delta(r)}dt^{2}+\frac{dr^{2}}{N(r)}+r^{2}\left(  d\theta
^{2}+\sin^{2}\theta d\varphi^{2}\right)  ,\qquad\mathbf{A}=A_{t}dt=V(r)dt,
\label{eq:metric ansatz}%
\end{equation}
we obtain the equations of motion
\begin{align}
2m^{\prime}(r)-r^{2}N(r)\phi^{\prime}(r)^{2}-e^{2\delta(r)+\alpha\phi(r)^{2}%
}r^{2}V^{\prime}(r)^{2}  &  =0,\nonumber\\
\delta^{\prime}(r)+r\phi^{\prime}(r)^{2}  &  =0,\nonumber\\
\left[  e^{-\delta(r)}r^{2}N(r)\phi^{\prime}(r)\right]  ^{\prime}-\alpha
e^{\delta(r)+\alpha\phi(r)^{2}}\phi(r)r^{2}V^{\prime}(r)^{2}  &
=0,\label{eq:eom}\\
\left[  e^{\delta(r)+\alpha\phi(r)^{2}}r^{2}V^{\prime}(r)\right]  ^{\prime}
&  =0,\nonumber
\end{align}
where $N(r)\equiv1-2m(r)/r$ can be expressed in terms of the Misner-Sharp mass
function $m(r)$, and primes denote derivatives with respect to $r$. The last
equation in Eq. $\left(  \ref{eq:eom}\right)  $ yields $V^{\prime
}(r)=-e^{-\delta(r)-\alpha\phi(r)^{2}}Q/r^{2}$, where the constant $Q$ can be
interpreted as the electric charge of the black hole. To solve the above
non-linear ODEs, suitable boundary conditions at the event horizon of radius
$r_{h}$ and spatial infinity shall be imposed as
\begin{align}
m(r_{h})  &  =\frac{r_{h}}{2},\qquad\delta(r_{h})=\delta_{0},\qquad\phi
(r_{h})=\phi_{0},\qquad V(r_{h})=0,\nonumber\\
m(\infty)  &  =M,\qquad\delta(\infty)=0,\qquad\phi(\infty)=0,\qquad
V(\infty)=\Psi,
\end{align}
where $\delta_{0}$ and $\phi_{0}$ are two positive constants, $M$ is the ADM
mass, and $\Psi$ is the electrostatic potential. The scalar-free black hole
solution with $\phi=0$ corresponds to Reissner-Nordstr\"{o}m black holes
(RNBHs). When the dimensionless coupling $\alpha$ is larger than $1/4$, there
exist HBH solutions with a non-trivial profile of the scalar field $\phi(r)$
\cite{Herdeiro:2018wub,Fernandes:2019rez,Wang:2020ohb}. In this paper, we
focus on the fundamental state of the HBH solutions, which means that
$\phi(r)$ remains positive outside the event horizon.

The behavior of a photon traveling outside the HBH can be encapsulated in the
null geodesic equation,
\begin{equation}
\frac{d^{2}x^{\mu}}{d\lambda^{2}}+\Gamma_{\rho\sigma}^{\mu}\frac{dx^{\rho}%
}{d\lambda}\frac{dx^{\sigma}}{d\lambda}=0, \label{eq:geodesic}%
\end{equation}
where $\lambda$ is the affine parameter, and $\Gamma_{\rho\sigma}^{\mu}$ is
the Christoffel symbol. In the appendix, we show that light rays propagating
in the HBH spacetime indeed \ are determined by the null geodesic equation.
Due to the spherical symmetry, we only consider light rays moving on the
equatorial plane with $\theta=\pi/2$. Combining $ds^{2}=0$ and Eqs. $\left(
\ref{eq:metric ansatz}\right)  $ and $\left(  \ref{eq:geodesic}\right)  $, one
obtains the time, azimuthal and radial components of the null geodesic,
\begin{align}
\frac{dt}{d\lambda}  &  =\frac{1}{bN(r)e^{-2\delta(r)}},\nonumber\\
\frac{d\varphi}{d\lambda}  &  =\pm\frac{1}{r^{2}},\label{eq:light ray eom}\\
e^{-2\delta(r)}\left(  \frac{dr}{d\lambda}\right)  ^{2}  &  =\frac{1}{b^{2}%
}-\frac{e^{-2\delta(r)}N(r)}{r^{2}},\nonumber
\end{align}
where $\pm$ in the second line corresponds to the light rays moving in the
counterclockwise $(+)$ or clockwise $(-)$ along $\varphi$-direction. The
impact parameter $b$ is defined as $|L|/E$, where $L$ and $E$ are the
conserved angular momentum and energy of the photons, respectively. Note that
we use a redefined affine parameter $\lambda\rightarrow\lambda/|L|$ in Eq.
$\left(  \ref{eq:light ray eom}\right)  $. From the last equation of Eq.
$\left(  \ref{eq:light ray eom}\right)  $, one can define the effective
potential of light rays as
\begin{equation}
V_{\text{eff}}(r)=\frac{e^{-2\delta(r)}N(r)}{r^{2}}. \label{eq:Veff}%
\end{equation}
Particularly, an unstable photon sphere (or equivalently, a circular and
unstable null geodesic) is determined by
\begin{equation}
V_{\text{eff}}(r_{ph})=\frac{1}{b_{ph}^{2}},\qquad V_{\text{eff}}^{\prime
}(r_{ph})=0,\qquad V_{\text{eff}}^{\prime\prime}(r_{ph})<0,
\label{eq:photon sphere}%
\end{equation}
where $r_{ph}$ is the radius of the photon sphere, and $b_{ph}$ is the
corresponding impact parameter. In this paper, we focus on unstable photon
spheres since they play an important role in determining the accretion disk
image seen by a distant observer. For convenience, photon spheres are referred
to unstable photon spheres in the remainder of this paper unless we make an
explicit statement.

In this paper, we consider that the HBH is surrounded by a static and
geometrically thin accretion disk, which is assumed to radiate isotropically
in the rest frame of the matter. In addition, we take the disk emission to be
optically thin by neglecting the absorption effect. Note that there exist
compelling evidences indicating that an optically thin hot accretion flow may
surround M87{*} or Sgr A{*} \cite{Yuan:2014gma,Johnson:2015iwg}. To obtain the
accretion disk image perceived by a distant observer, we evolve light rays
from the observer's position backwards in time. Furthermore, one can use $m$,
the times that a certain ray intersects with the disk plane outside the
horizon, to distinguish light rays' behavior. Following the definitions in
\cite{Gralla:2019xty}, light rays with $m\leq1$, $m=2$ and $m\geq3$ constitute
the direct emission, the lensing ring and the photon ring, respectively. The
observed total intensity $F_{o}(b)$ generated by a light ray of impact
parameter $b$ is a sum of the intensities from each intersection with the disk
plane outside the horizon,
\begin{equation}
F_{o}(b)=\int_{\nu_{o}}d\nu_{o}I_{\nu_{o}}(b)=\underset{m}{\sum}\left.
\int_{\nu_{e}}g(r)d\nu_{e}g^{3}(r)I_{\nu_{e}}(r)\right\vert _{r=r_{m}%
(b)}=\underset{m}{\sum}\left.  g^{4}(r)F_{e}(r)\right\vert _{r=r_{m}(b)},
\label{eq:intensity}%
\end{equation}
where $g(r)$ is the red-shift factor, $I_{\nu_{e}}$ is the specific intensity
at the emission frequency $\nu_{e}$, $I_{\nu_{o}}$ is the specific intensity
at the observed frequency $\nu_{o}$, and $F_{e}=\int_{\nu_{e}}I_{\nu_{e}}%
d\nu_{e}$ is the total emitted intensity. Here, we have applied the
$I_{\nu_{o}}=g^{3}(r)I_{\nu_{e}}$ by Liouville's theorem in the second
equality. The function $r_{m}(b)$ $(m=1,2,3...)$, dubbed the transfer
function, is the radial coordinate of the light ray crossing the disk plane at
the $m^{th}$ time. Moreover, the slope of the transfer function, $dr_{m}/db$,
is the demagnification factor of the $m^{th}$ image of the accretion disk. For
simplicity, we set the total emitted intensity $F_{e}(r)=1/r^{2}$ with
$r>r_{h}$, which suffices to illustrate the major features of the accretion
disk image. In this case, the observed total intensity is given by
\begin{equation}
F_{o}(b)=\underset{m}{\sum}\left.  \frac{N^{2}(r)e^{-4\delta(r)}}{r^{2}%
}\right\vert _{r=r_{m}(b)}, \label{eq:intensity2}%
\end{equation}
where we use $g(r)=\sqrt{N(r)}e^{-\delta(r)}$ \cite{Gan:2021pwu}. When viewed
in a face-on orientation, the 2D image of the accretion disk is circularly
symmetric. Taking account of the geometric interpretation of $b$, we can
depict the 2D face-on image by employing $F_{o}(b)=F_{o}(\sqrt{X^{2}+Y^{2}})$,
where the coordinates $(X,Y)$ span the observer's celestial plane.

\section{Numerical Results}

\label{sec:numerical result}

\begin{figure}[ptb]
\begin{centering}
\includegraphics[scale=0.46]{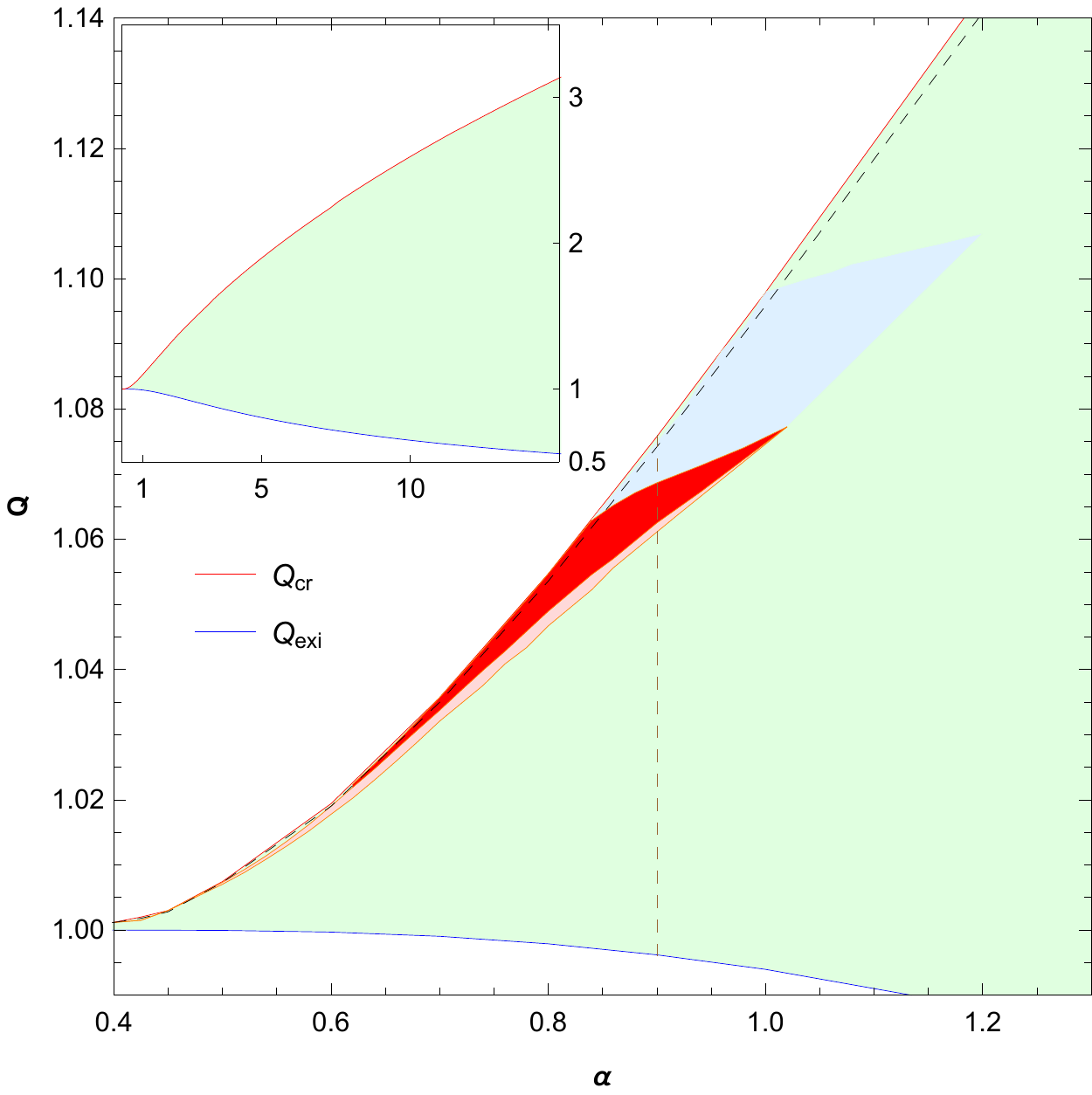}$\,$\includegraphics[scale=0.65]{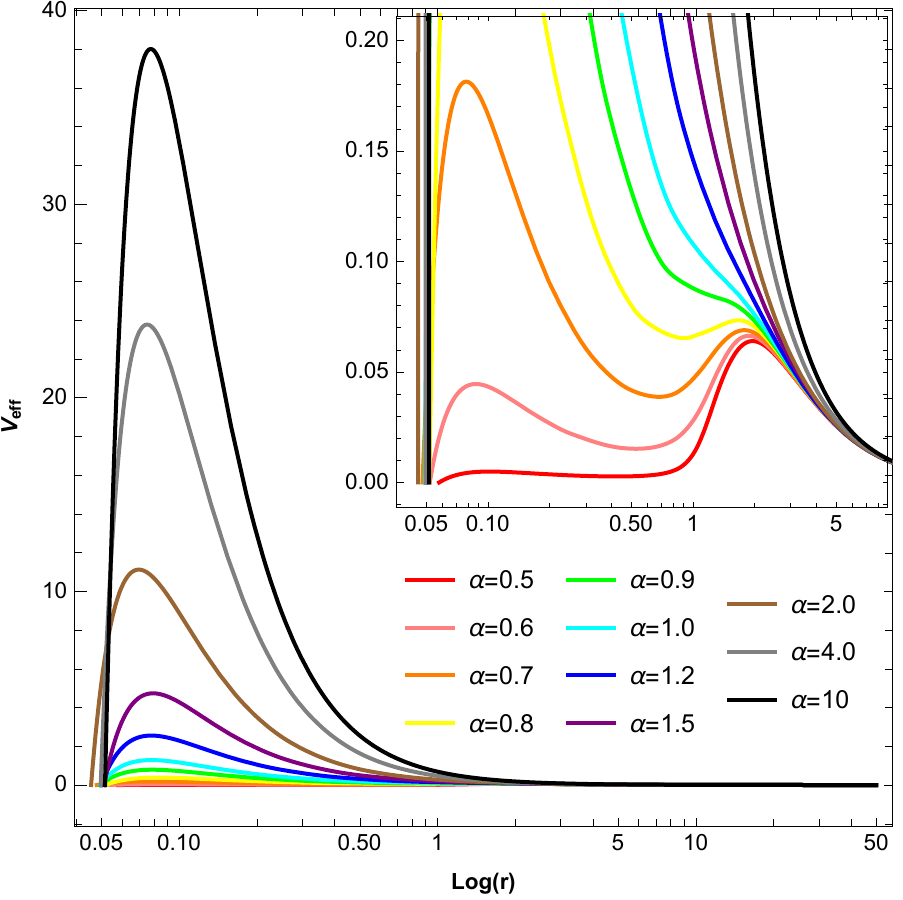}$\,$\includegraphics[scale=0.65]{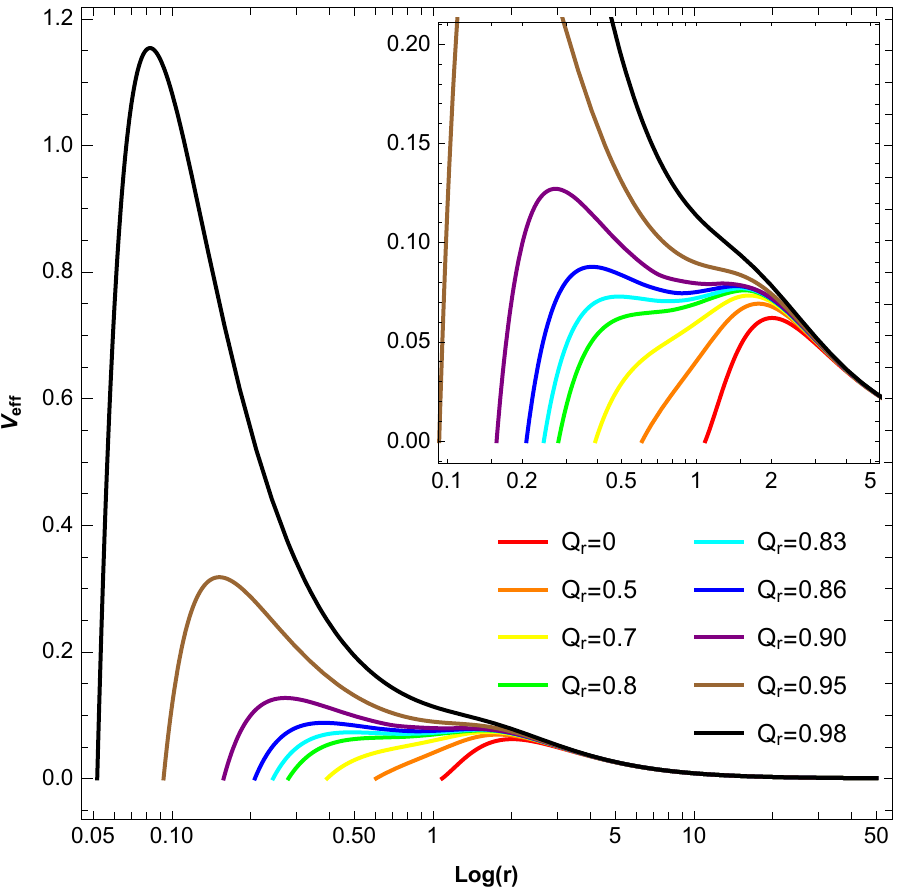}
\par\end{centering}
\caption{$\mathbf{Left}$: Regions of four different HBH families in the
$Q$-$\alpha$ plane: single-peak I (green), single-peak II (blue), double-peak
I (orange) and double peak II (red). The domain of the HBH solutions is
bounded by the existence line (blue) and the critical line (red). The black
dashed line and the vertical dashed line denote $Q_{r}\equiv(Q-Q_{\text{exi}%
})/(Q_{\text{cr}}-Q_{\text{exi}})=0.98$ and $\alpha=0.9$, respectively.
$\mathbf{Middle}$\textbf{:} The profiles of the effective potentials at fixed
$Q_{r}=0.98$ for different $\alpha$. $\mathbf{Right}$\textbf{:} The profiles
of the effective potentials at fixed $\alpha=0.9$ for different $Q_{r}$.}%
\label{figure veff}%
\end{figure}

In this section, we present the numerical results about the optical appearance
of\ the accretion disk surrounding the HBH black hole, viewed face-on. In the
left panel of Fig. \ref{figure veff}, we first display the domain of existence
for HBH solutions to the action $\left(  \ref{eq:action}\right)  $, which is
bounded by the sets of existence and critical solutions, denoted by
$Q_{\text{exi}}$-line (red line) and $Q_{\text{cr}}$-line (blue line) in the
$Q$-$\alpha$ plane, respectively. On the $Q_{\text{cr}}$-line, the black hole
horizon radius vanishes with the black hole mass and charge remaining finite.
Furthermore, this domain can be divided into four parameter space regions, in
one of which the effective potentials $\left(  \ref{eq:Veff}\right)  $ have
distinct profiles, e.g., the number of the maxima. In particular, we obtain
four families of the HBH solutions,

\begin{enumerate}
\item Single-peak I (green region): The potential has a single maximum, which
is similar to Schwarzschild and RN black holes
\cite{Eiroa:2002mk,Narayan:2019imo}. For instance, the $\alpha=10$ case (black
line) in the middle panel of Fig. \ref{figure veff}.

\item Single-peak II (blue region): The potential has a single maximum and an
\textquotedblleft ankle-like\textquotedblright\ structure, e.g., the
$\alpha=0.9$ case (green line) in the middle panel. The ankle-like structure
is characterized by a flattening of the potential, and corresponds to a
transition between convexity and concavity of the effective potential. More
precisely, its presence can be determined by the appearance of $V_{\text{eff}%
}^{\prime\prime}(r)>0$ in the region where $V_{\text{eff}}^{\prime}(r)<0$.

\item Double-peak I (orange region): The potential has two maxima at two
different radii, and the peak of the potential at the smaller radius is lower
than that at the larger radius, e.g., the $\alpha=0.6$ case (pink line) in the
middle panel.

\item Double-peak II (red region): The potential has two maxima at two
different radii, and the peak of the potential at the smaller radius is higher
than that at the larger radius, e.g., the $\alpha=0.7$ case (orange line) in
the middle panel.
\end{enumerate}

It is observed that the single-peak I family occupies almost the whole HBH
existence regime, whereas the other three families only exist in a small
$\alpha$ region near the $Q_{\text{cr}}$-line.

To illustrate how the potential profile changes among different families, we
display a set of potentials along the lines with fixed $Q_{r}\equiv
(Q-Q_{\text{exi}})/(Q_{\text{cr}}-Q_{\text{exi}})=0.98$ and fixed $\alpha=0.9$
in the middle and right panels of Fig. \ref{figure veff}, respectively. As
$\alpha$ increases with $Q_{r}=0.98$, the middle panel (specifically, the
inset therein) shows that the potential first has a single peak, then another
peak at a smaller radius\ appears and grows, meanwhile the peak at the larger
radius shrinks until disappears, indicating the single-peak I $\rightarrow$
double-peak I $\rightarrow$ double-peak II $\rightarrow$ single-peak II
$\rightarrow$ single-peak I transition. When $Q_{r}$ increases with fixed
$\alpha=0.9$, the right panel and the inset therein present the single-peak I
$\rightarrow$ double-peak I $\rightarrow$ double-peak II $\rightarrow$
single-peak II transition. In what follows, we display several representative
cases to show the main features of each family. Note that the HBH mass $M$ is
set to $1$ without loss of generality in the rest of this section.

\subsection{Single-peak potential}

\label{sec:One peak}

\begin{figure}[ptb]
\begin{centering}
\includegraphics[scale=0.45]{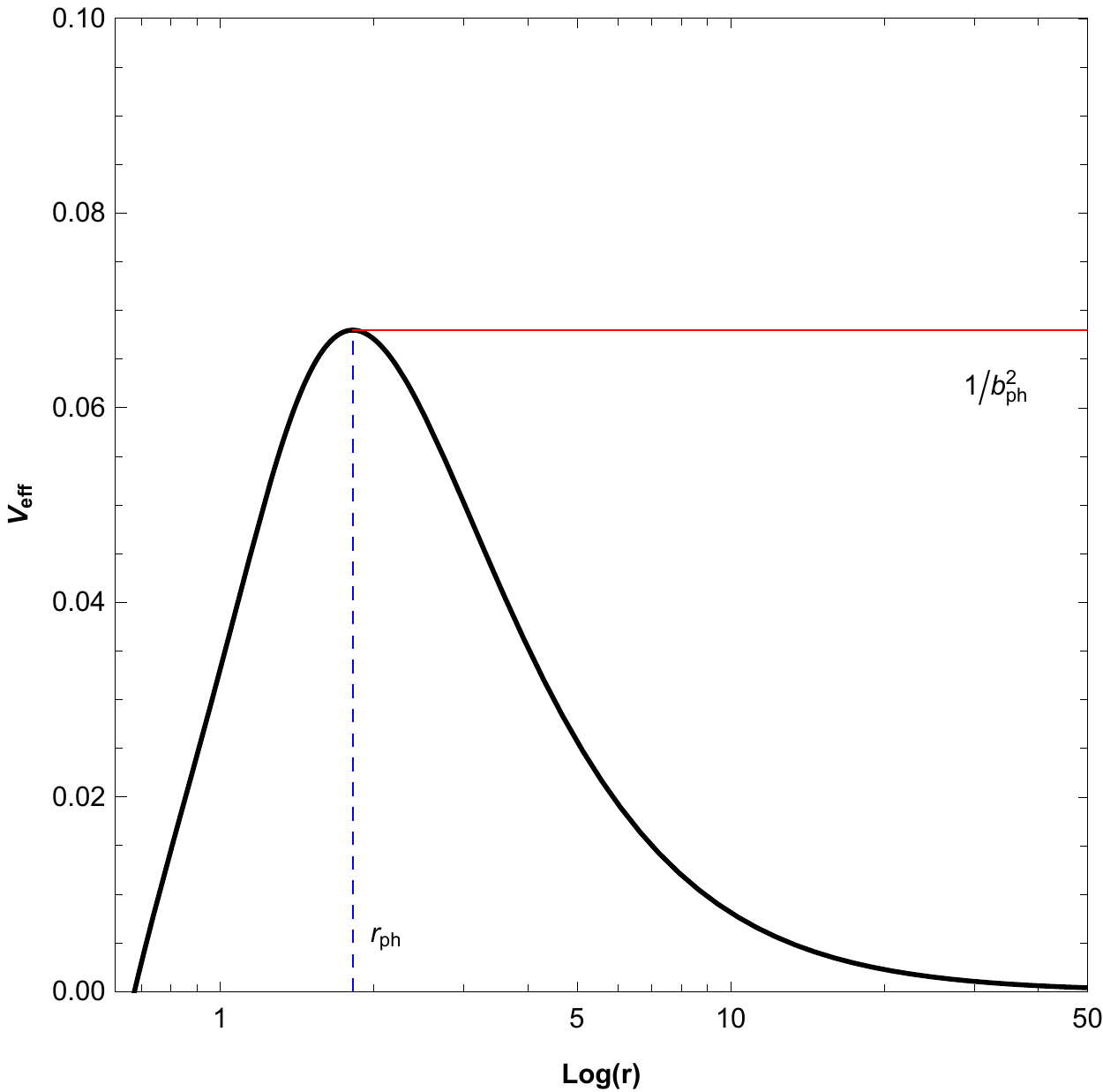}$\,$\includegraphics[scale=0.6]{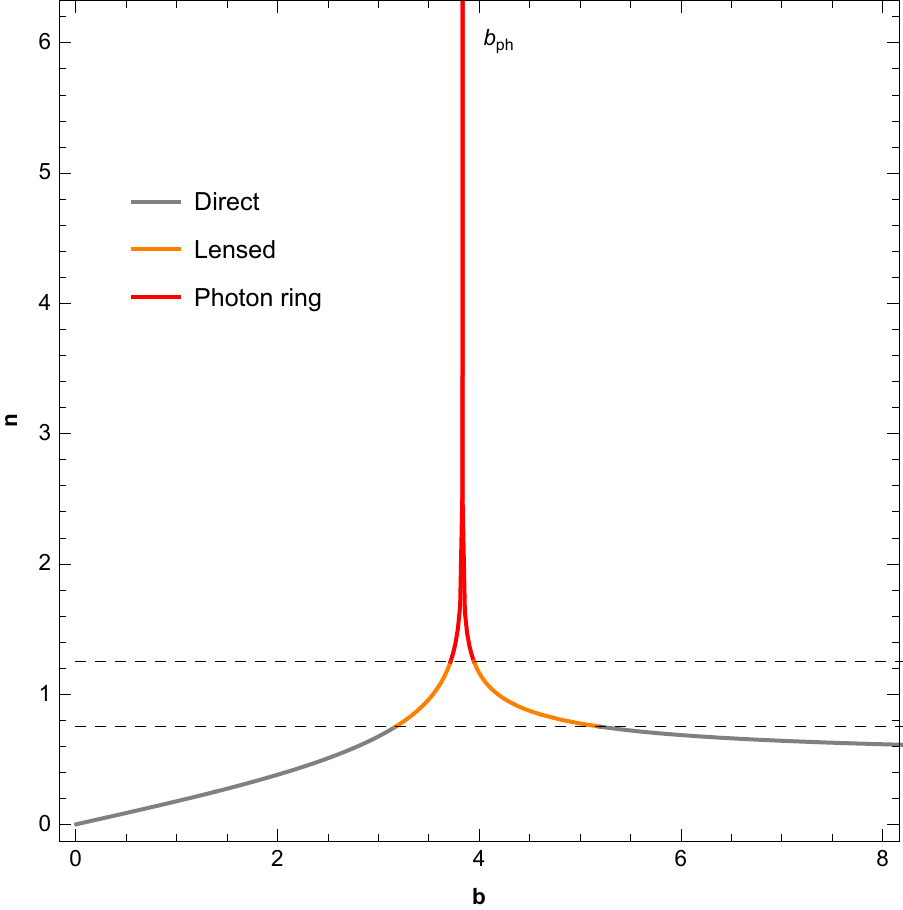}$\,$\includegraphics[scale=0.61]{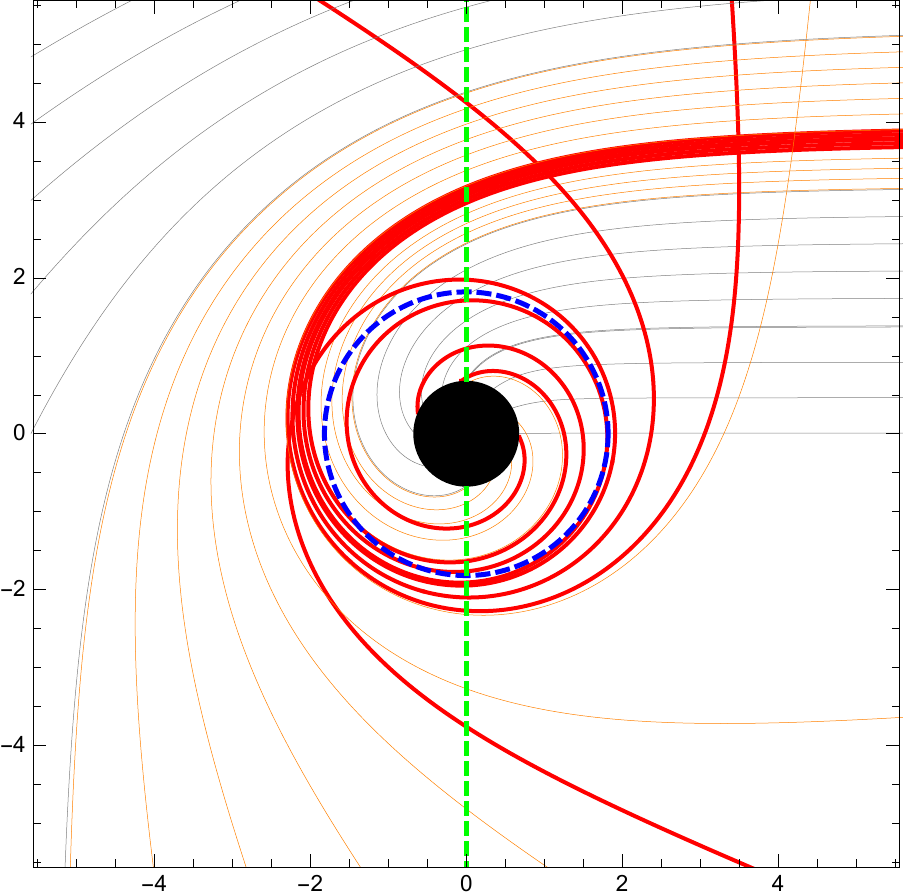}
\par\end{centering}
\vspace{5mm} \begin{centering}
\includegraphics[scale=0.42]{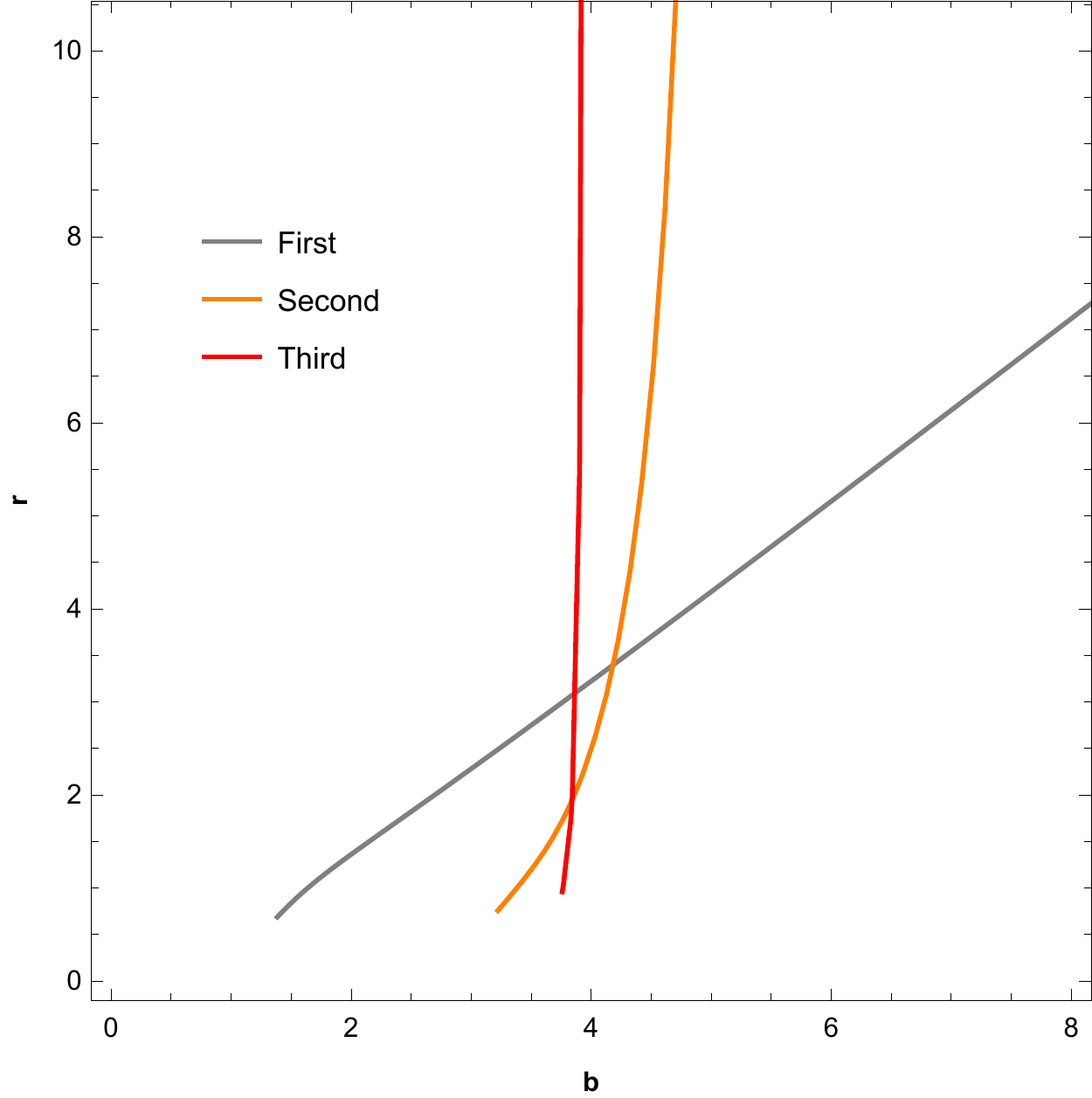}$\,$\includegraphics[scale=0.43]{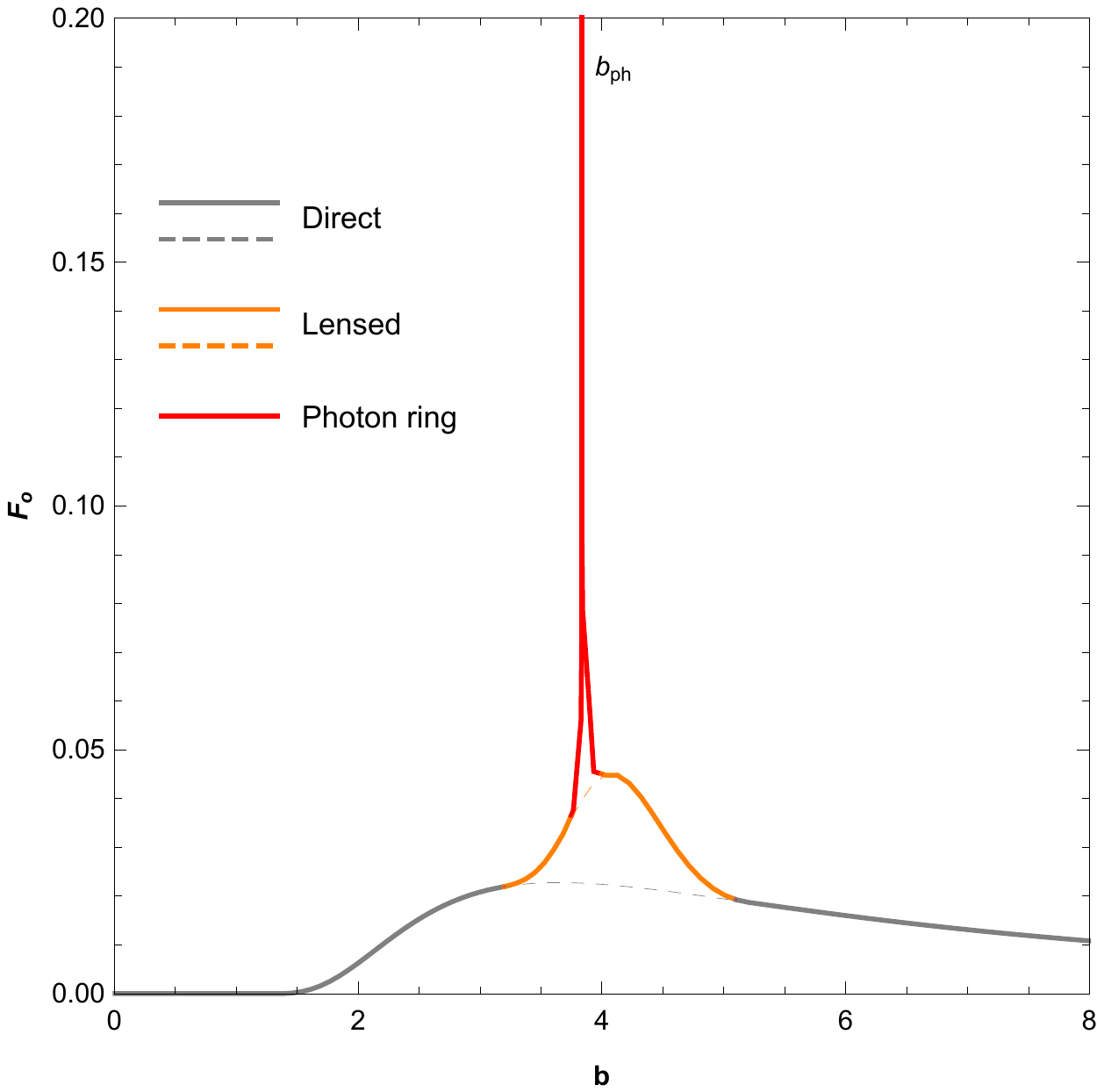}$\,$\includegraphics[scale=0.585]{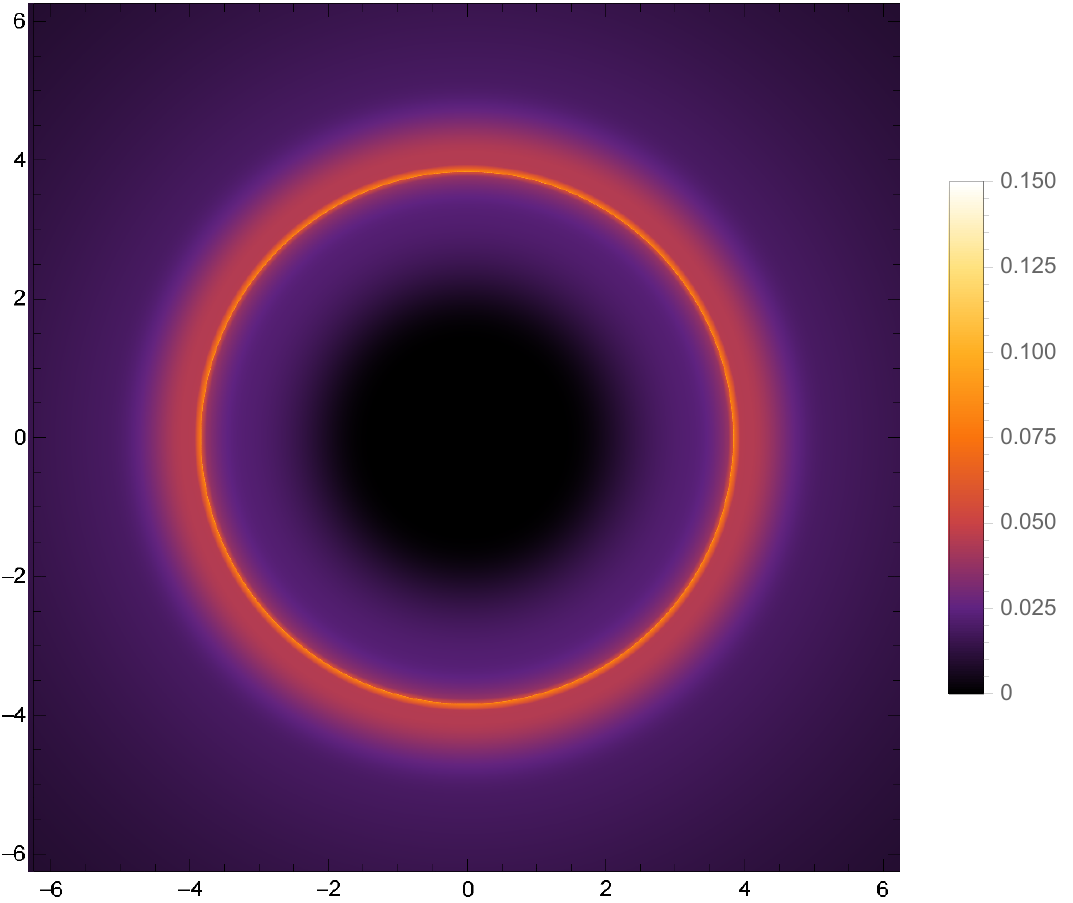}
\par\end{centering}
\caption{ Behavior of photons in a HBH with $\alpha=0.9$ and $Q=1.03$, which
is in the single-peak I family, and the observational appearance of an
optically and geometrically thin accretion disk around the HBH, viewed from a
face-on orientation. $\mathbf{Upper}$ $\mathbf{Left}$: The profile of the
effective potential, which possesses a single maximum. $\mathbf{Upper}$
$\mathbf{Middle}$: The total number of orbits $n=\Phi/(2\pi)$, where
$\Phi=\Delta\varphi$ is the total change of the azimuthal angle outside the
event horizon. The direct emission, the lensing ring and the photon ring
correspond to $n<0.75$ (gray), $0.75\leq n<1.25$ (orange), and $n\geq1.25$
(red), respectively. $\mathbf{Upper}$ $\mathbf{Right}$: A selection of photon
trajectories within the direct, lensed and photon ring bands. The blue dashed
circle is the photon sphere, and the green line denotes the cross section of
the disk plane. $\mathbf{Lower}$ $\mathbf{Left}$: The first three transfer
functions $r_{m}(b)$ with $m=1$ (gray), $m=2$ (orange) and $m=3$ (red),
representing the radial coordinates of the first, second and third
intersections with the accretion disk. $\mathbf{Lower}$ $\mathbf{Middle}$: The
observed total intensity $F_{o}(b)$ and the net contributions to the total
flux from the direct (gray), lensed (orange) and photon ring (red) bands.
$\mathbf{Lower}$ $\mathbf{Right}$: The 2D image of the accretion disk viewed
in the observer's sky. A bright and narrow ring appears at the location of the
photon ring. }%
\label{figure single peak 1}%
\end{figure}

For the single-peak I family, we consider a HBH with $\alpha=0.9$ and $Q=1.03$
in Fig. \ref{figure single peak 1}. The potential with one maximum indicates a
single photon sphere with radius $r_{ph}$ and the corresponding impact
parameter $b_{ph}$. To consider the accretion disk viewed from a face-on
orientation, we assume that an observer is placed at the far right of the
upper-right panel, corresponding to the \textquotedblleft north pole
direction\textquotedblright, and the disk lies in the equatorial plane with
respect to the observer's orientation (dashed green line in the upper-right
panel). Tracing a light ray backwards from the observer, the total orbit
number $n$ is related to the total disk-crossing times $m$ as $n<0.5$
$\rightarrow m=0$, $0.5\leq n<0.75$ $\rightarrow m=1$, $0.75\leq n<1.25$
$\rightarrow m=2$ and $n\geq1.25$ $\rightarrow m\geq3$. By definition,
$m\leq1$, $m=2$ and $m\geq3$ correspond to the direct emission, the lensing
ring and the photon ring, respectively \cite{Gralla:2019xty}. In the
upper-middle panel, we plot $n$ as a function of $b$ and depict the direct,
lensing and photon ring bands in gray, orange and red, respectively.

To obtain the observed intensity, we first calculate the transfer functions
$r_{m}(b)$ with $m=1,2,3,\cdots$. In the lower-left panel of Fig.
\ref{figure single peak 1}, we depict the transfer functions for $m=1,2$ and
$3$, which are associated with the direct, lensing and photon ring bands,
respectively. Since the average slope of $r_{m}(b)$ roughly reflects the
demagnified level of the $m^{th}$ image of the disk plane, it shows that the
secondary image is highly demagnified, and the tertiary image is extremely
demagnified. Via Eq. $\left(  \ref{eq:intensity2}\right)  $, the observed
intensity $F_{o}$ as a function of $b$ is shown in the lower-middle panel,
which presents a narrow spike-like photon ring (red) and a broader bump-like
lensing ring (orange) superimposed on the direct emission (gray). One can see
that the direct emission makes the dominant contribution to the overall
intensity flux, whereas the lensing%
$\backslash$%
photon ring makes modest%
$\backslash$%
little contributions. To present the 2D image of the accretion disk seen by a
distant observer, we project $F_{o}(b)$ to the observer's celestial
$(X,Y)$-plane via $b^{2}=X^{2}+Y^{2}$. In the lower-right panel, the
observational appearance of the photon ring is shown to be a thin bright ring
of radius $b_{ph}$, which suggests that the photon ring makes a very small
contribution to the total flux. Moreover, there exists a completely dark area
with vanishing intensity inside the photon ring, which is determined by the
$m=0$ band.

\begin{figure}[ptb]
\begin{centering}
\includegraphics[scale=0.45]{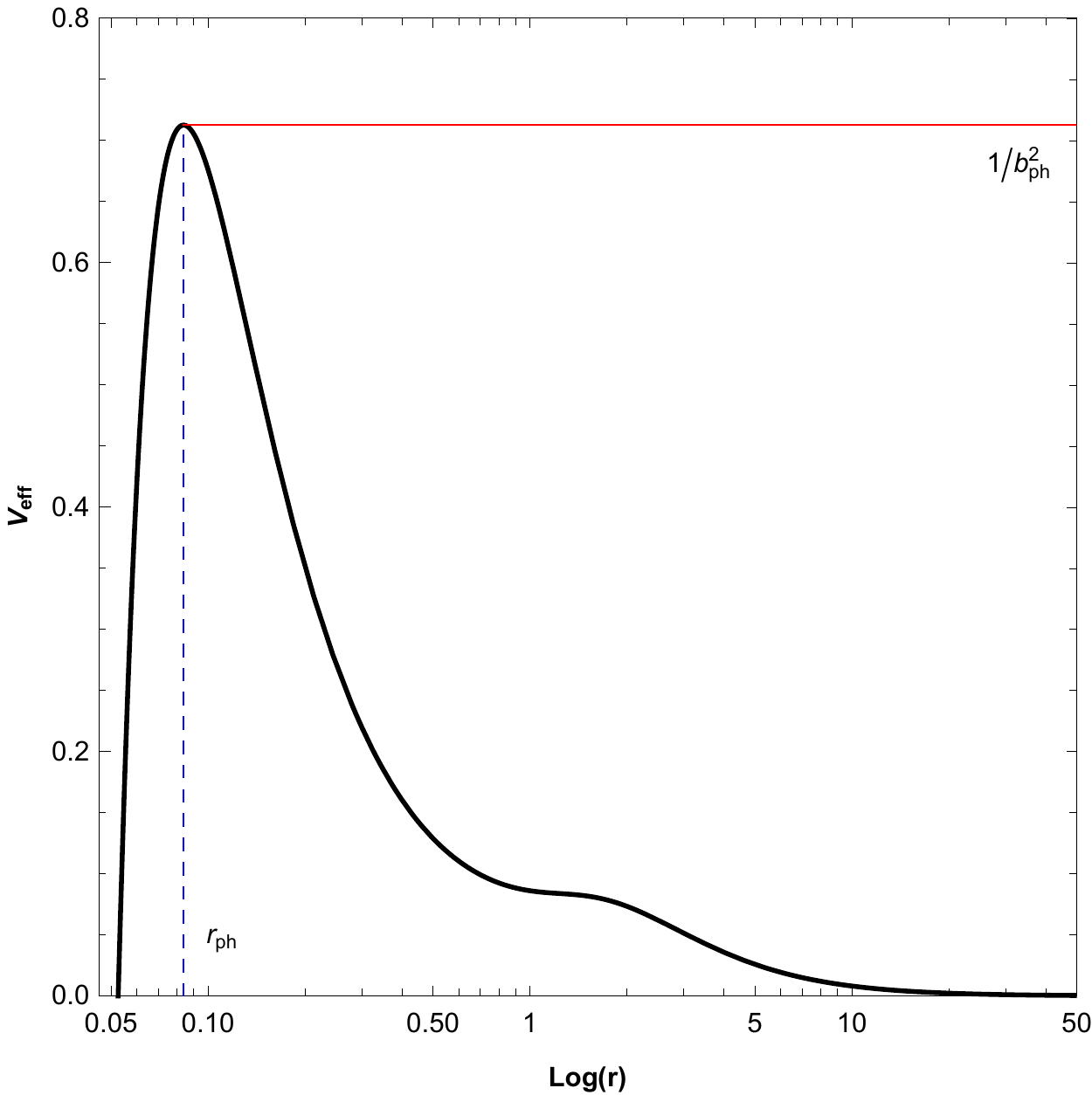}$\,$\includegraphics[scale=0.6]{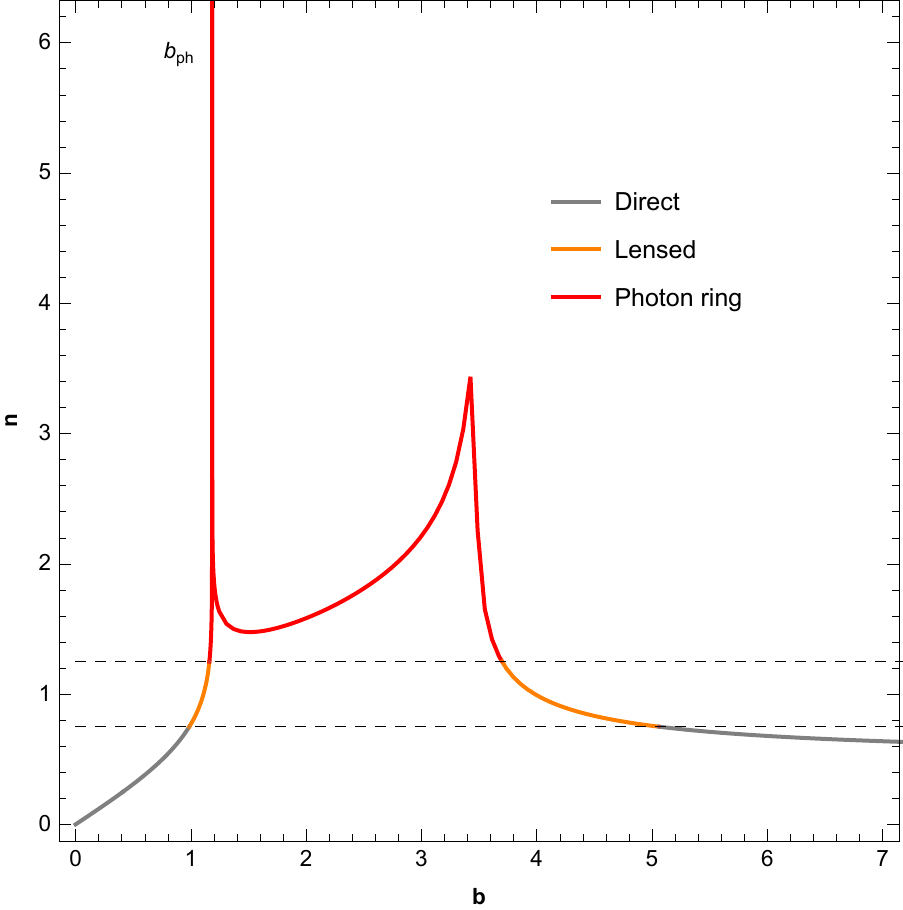}$\,$\includegraphics[scale=0.61]{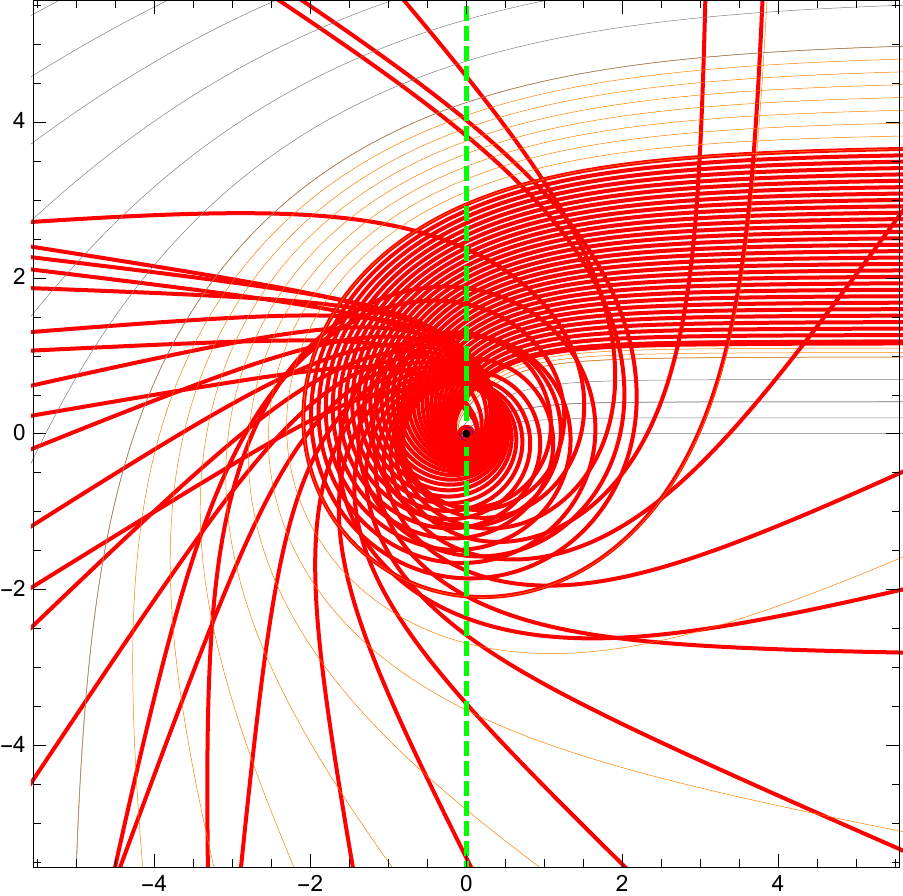}
\par\end{centering}
\vspace{5mm} \begin{centering}
\includegraphics[scale=0.42]{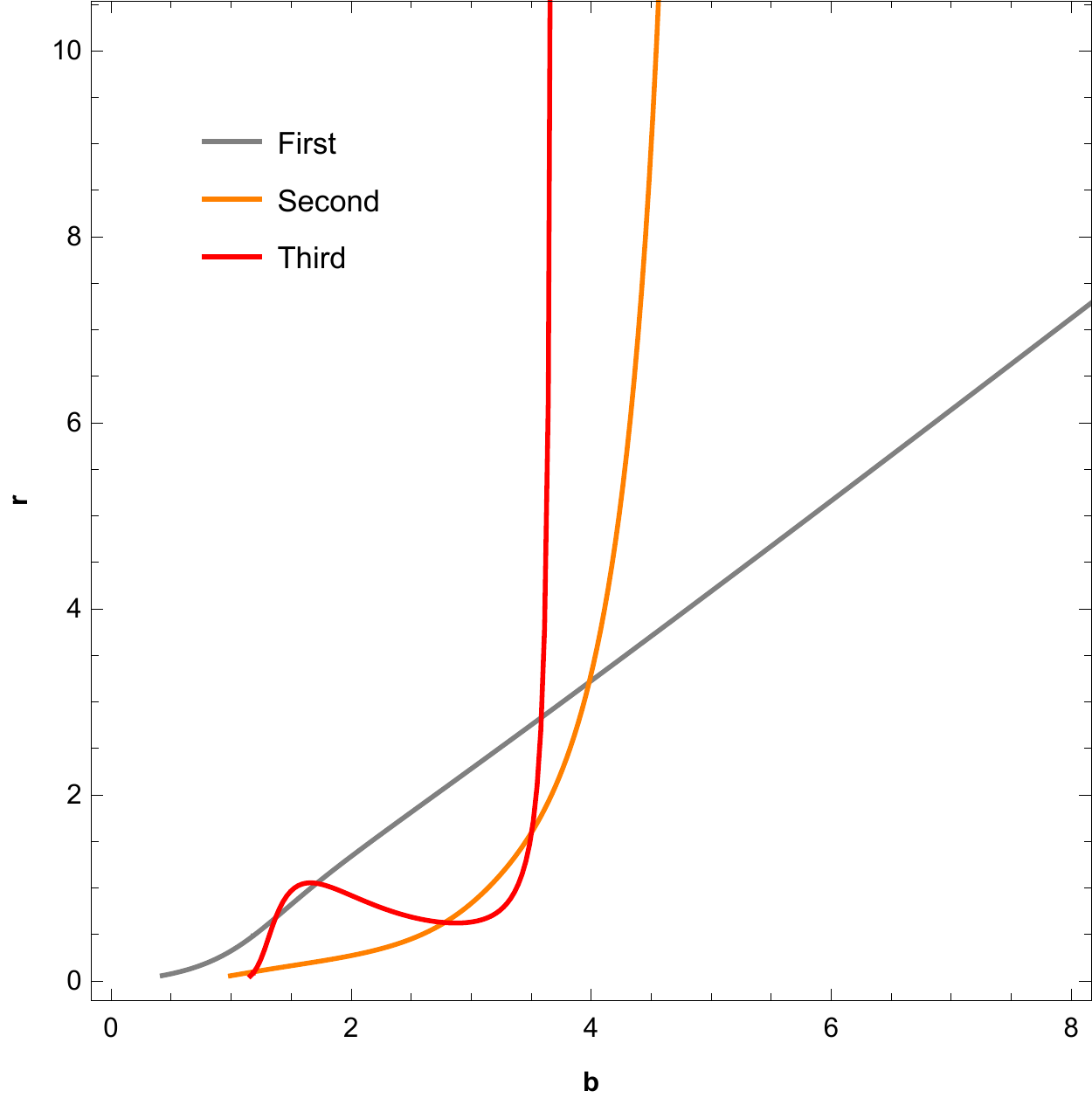}$\,$\includegraphics[scale=0.43]{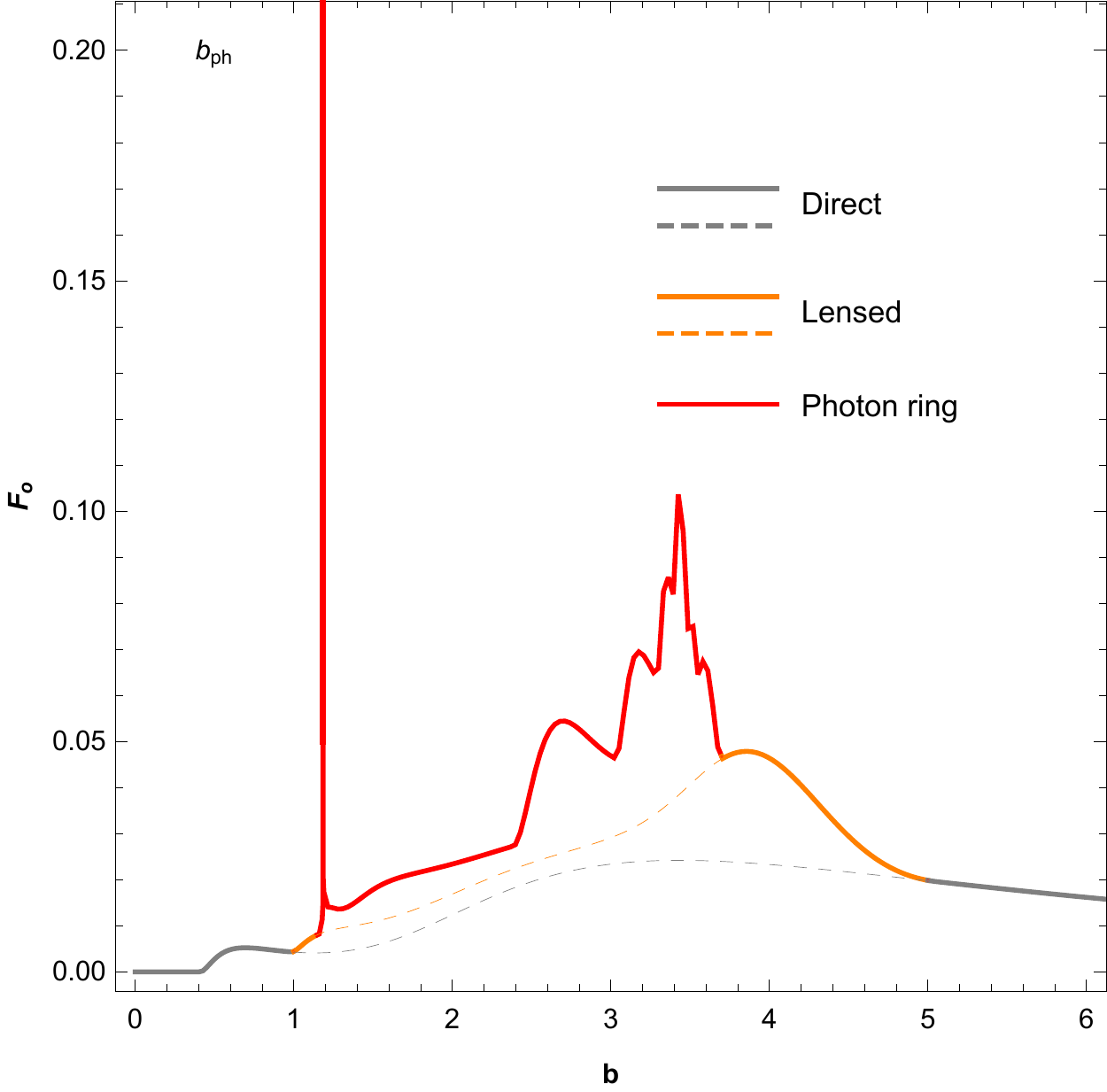}$\,$\includegraphics[scale=0.585]{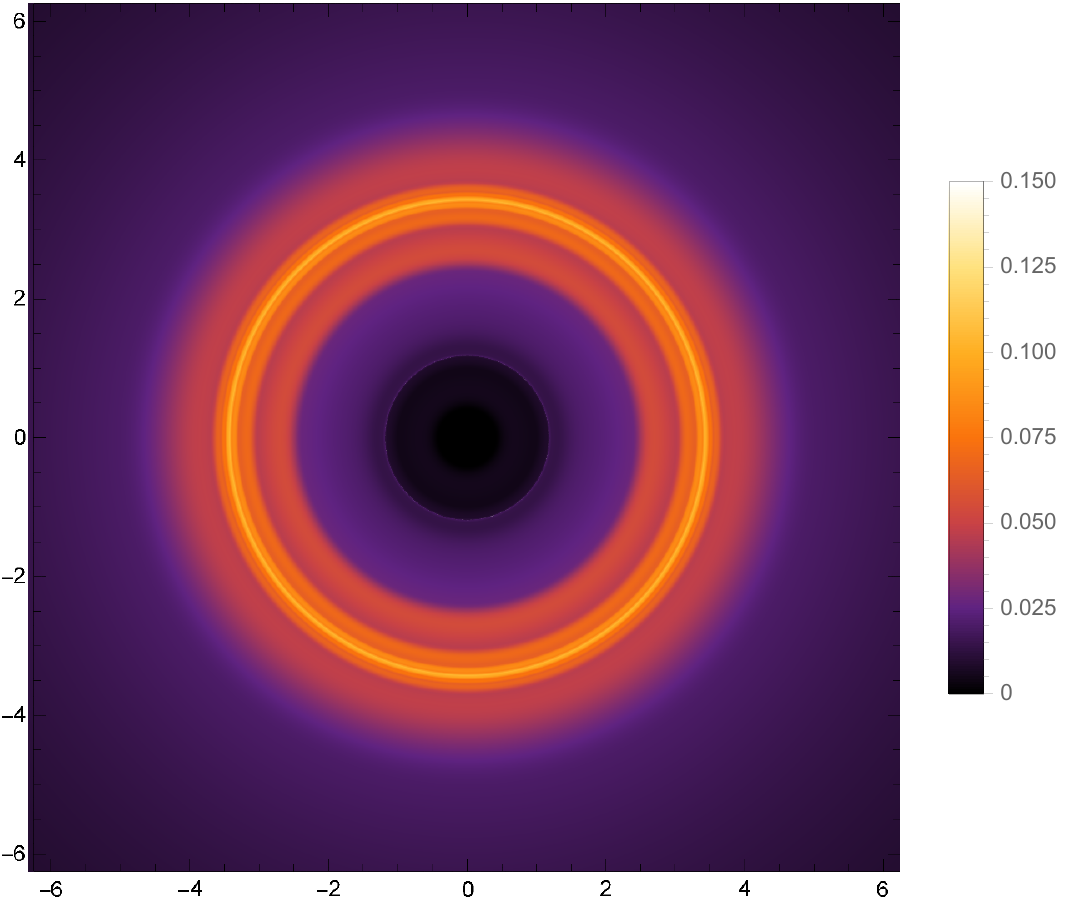}
\par\end{centering}
\caption{Behavior of photons in a HBH with $\alpha=0.9$ and $Q=1.074$, which
is in the single-peak II family, and the observational appearance of an
optically and geometrically thin accretion disk around the HBH, viewed from a
face-on orientation. The effective potential shows a flattening around $\log
r\sim1$ (\textbf{Upper Left}), which leads to a finite broad peak in the
$n$-$b$ plane (\textbf{Upper Middle}). As a result, the width of the photon
ring is significantly increased, and comparable to that of the lensing ring.
Therefore, the photon ring can make a non-trivial net contribution to the
observed flux $F_{o}(b)$ (\textbf{Lower Middle}). Specially, a bright and wide
annulus, consisting of multiple concentric bright thin rings, appears outside
the bright ring, corresponding to the photon sphere, in the accretion disk
image (\textbf{Lower Right}). Moreover, as indicated by the transfer
functions, the secondary and tertiary images are less demagnified than those
in the single-peak I family (\textbf{Lower Left}).}%
\label{figure single peak 2}%
\end{figure}

On the other hand, the photon ring can play a non-trivial role in the
observational appearance of an accretion disk around a HBH in the single-peak
II family. A HBH with $\alpha=0.9$ and $Q=1.074$ is considered in Fig.
\ref{figure single peak 2}, the upper-left panel of which shows that the
effective potential has an ankle-like structure around $\log r\sim1$. This
distinctive feature results in more complicated behavior of null geodesics and
observational appearance of emission from an accretion disk. Indeed, due to
the flattening of the ankle-like structure, photons can revolve around the
ankle-like structure multiple times, and cross the disk plane more than twice.
So the upper-middle panel displays that, in addition to the sharp peak
determined by the photon sphere at $b=b_{ph}$, the $n(b)$ curve also possesses
a much broader peak of finite height at $b>b_{ph}$, which significantly widens
the photon ring band. Note that a similar phenomenon has been reported in a
wormhole scenario \cite{Shaikh:2018oul}. Consequently, as shown in the
lower-middle pane, the photon ring makes a noticeable contribution to the
total intensity flux, comparable to that of the lensing ring. Therefore, the
photon ring plays an important role in determining the observational
appearance of the accretion disk, leading to the presence of a bright ring and
a concentric bright annular region, which can be observed in the lower-right
panel. While the bright ring at the smaller radius, corresponding to the
photon sphere, is barely visible due to the sharpness of the peak of
$F_{o}(b)$ at $b=b_{ph}$, the wide bright annular region at the larger radius
is quite noticeable and comprises multiple concentric thin bright rings with
different luminosity. Interestingly, the second and third transfer functions
both exist over a larger range of $b$ than those in the single-peak I family
(lower-left panels of Figs. \ref{figure single peak 1} and
\ref{figure single peak 2}), thus resulting in less demagnification for
secondary and tertiary images in the single-peak II family.

\subsection{Double-peak potential}

\label{sec:double peak}

\begin{figure}[ptb]
\begin{centering}
\includegraphics[scale=0.45]{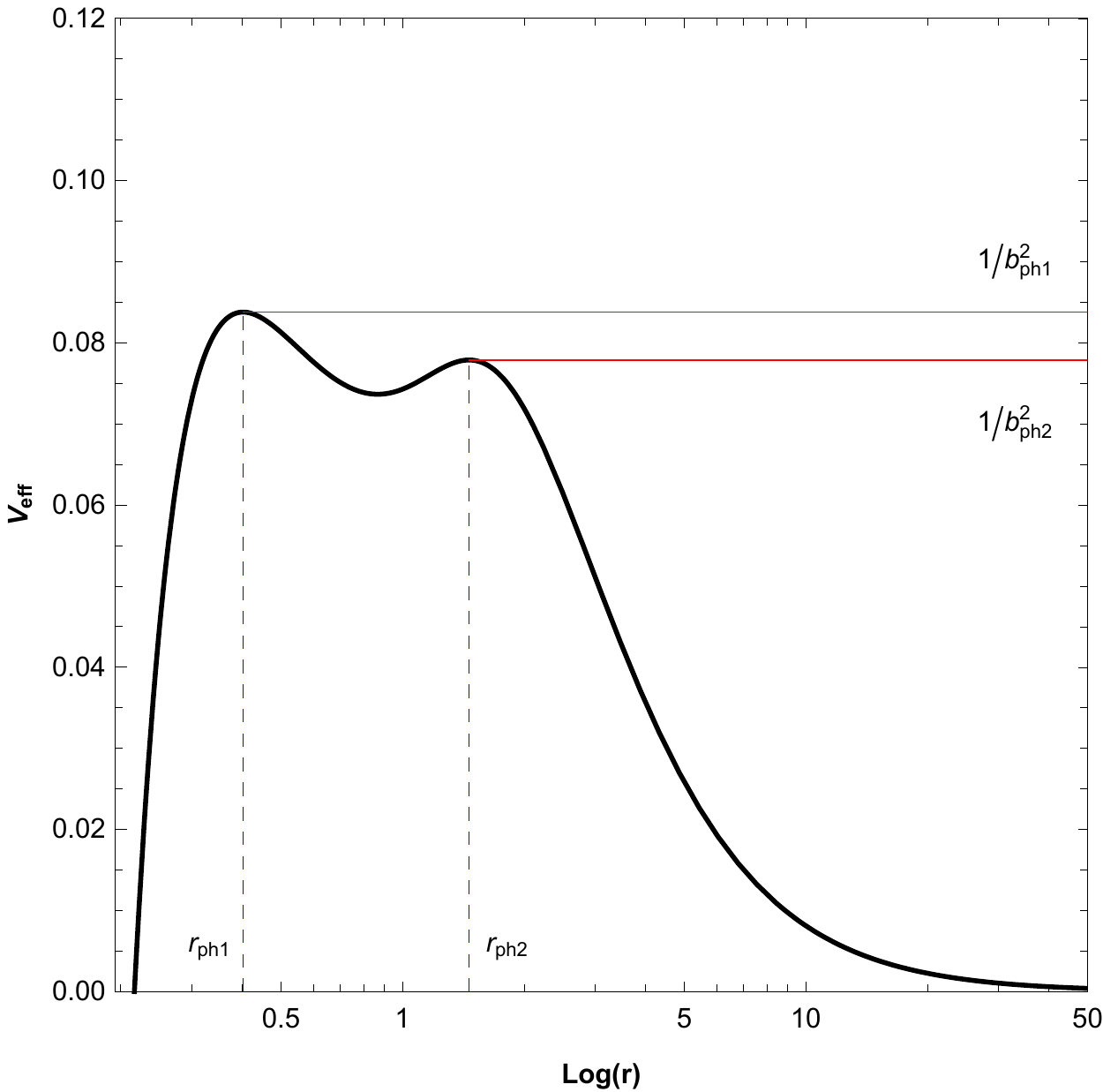}$\,$\includegraphics[scale=0.6]{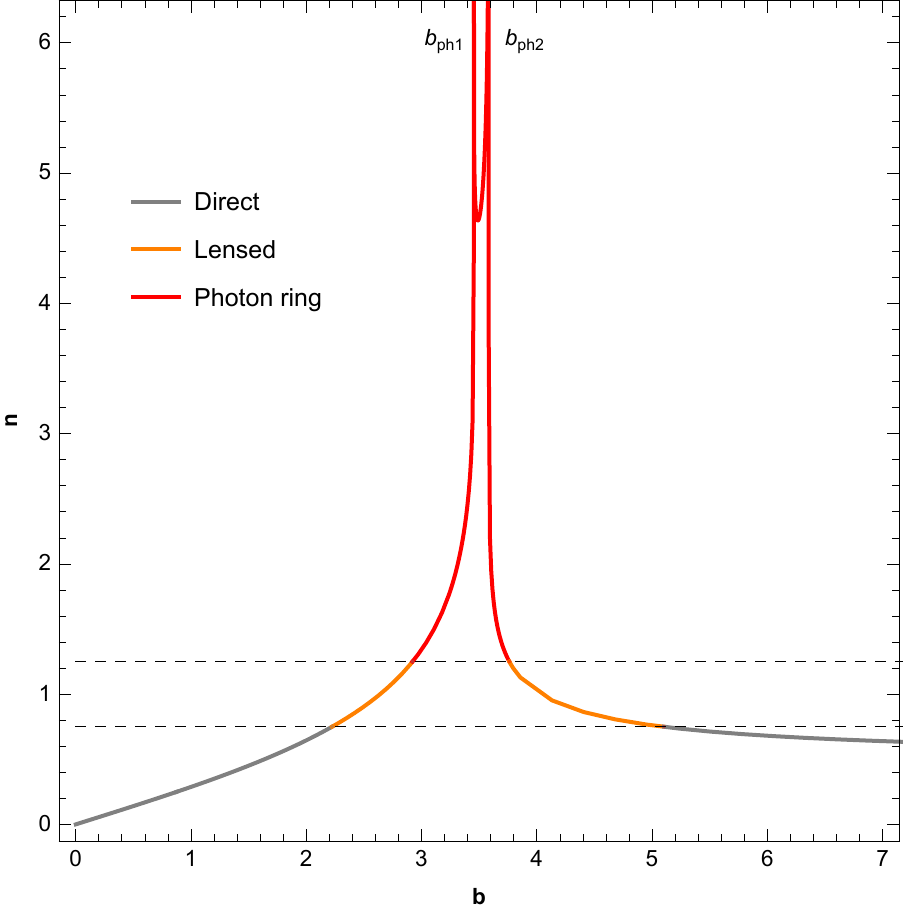}$\,$\includegraphics[scale=0.61]{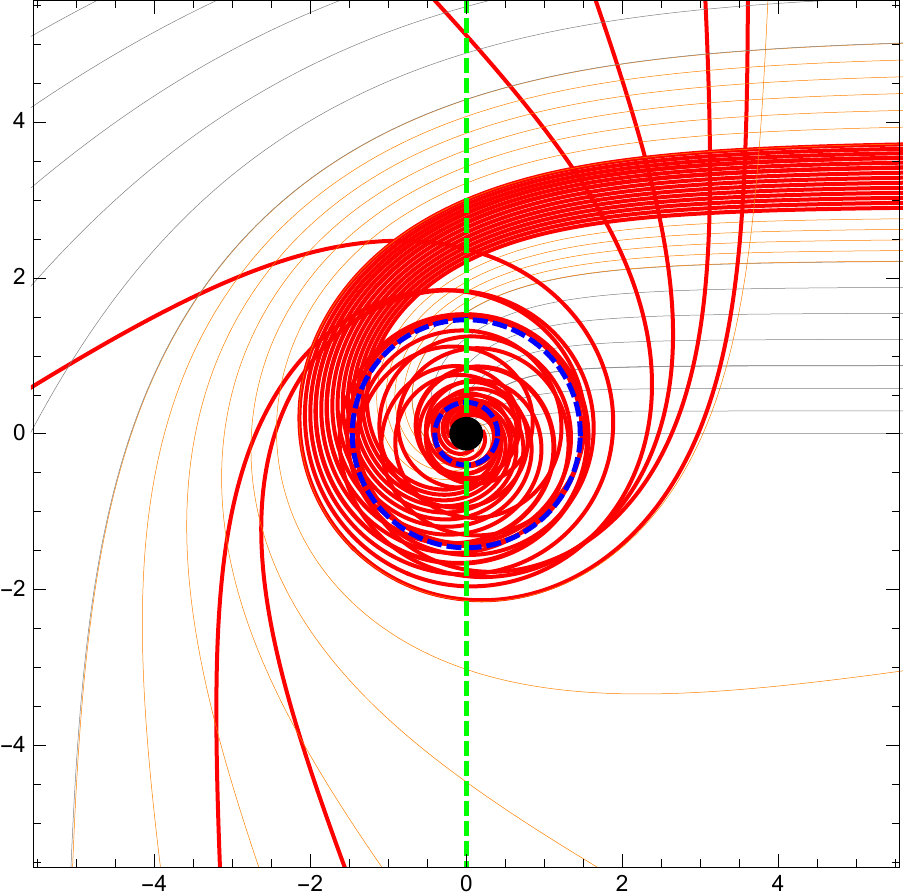}
\par\end{centering}
\vspace{5mm} \begin{centering}
\includegraphics[scale=0.42]{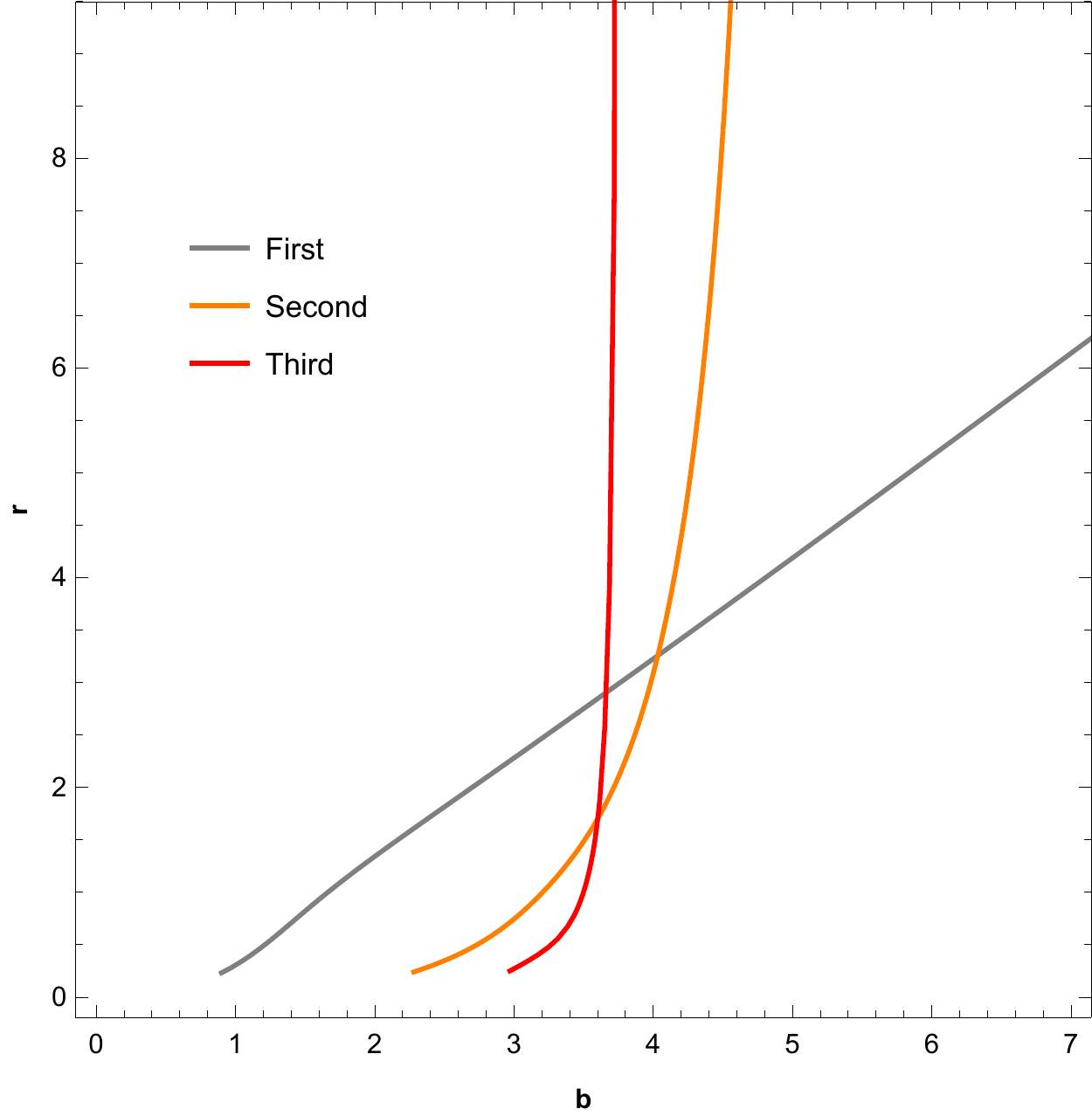}$\,$\includegraphics[scale=0.43]{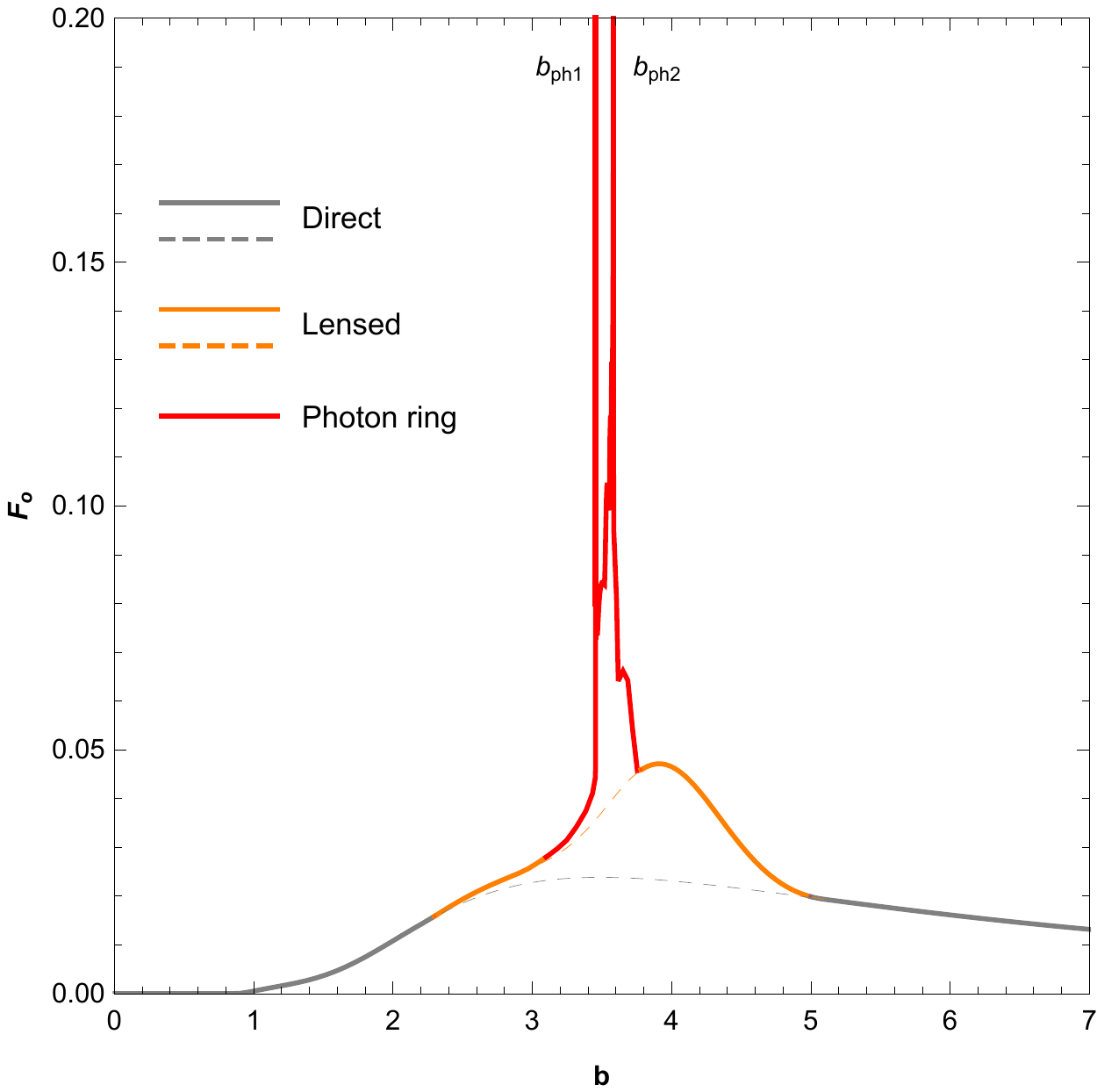}$\,$\includegraphics[scale=0.585]{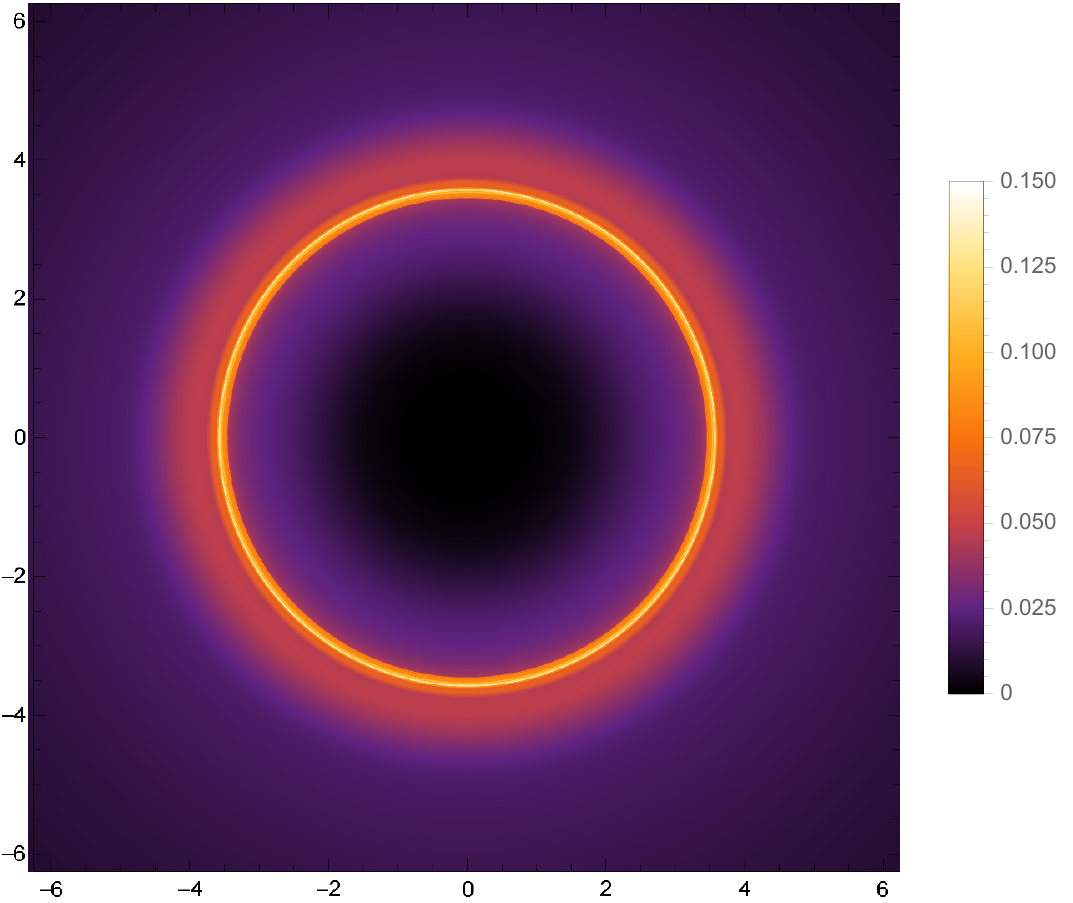}
\par\end{centering}
\caption{Behavior of photons in a HBH with $\alpha=0.9$ and $Q=1.064$, which
is in the double-peak II family, and the observational appearance of an
optically and geometrically thin accretion disk around the HBH, viewed from a
face-on orientation. The upper-left panel shows that the effective potential
has two peaks of similar heights at $r=r_{ph1}$ and $r=r_{ph1}$, respectively.
These two peaks correspond to two photon spheres (blue dashed circles),
leading to two infinite peaks in $n(b)$ and $F_{o}(b)$ curves and a narrow
photon ring (\textbf{Upper Middle and Lower Middle}). Due to the two photon
spheres, the contribution from the photon ring to the observed flux $F_{o}(b)$
is comparable to that of the lensing ring, and hence is non-negligible
(\textbf{Lower Middle}). However, the internal structure of the photon ring is
hardly visible in the accretion disk image because of the narrowness of the
photon ring (\textbf{Lower Right}).}%
\label{figure double peak 1}%
\end{figure}

\begin{figure}[ptb]
\begin{centering}
\includegraphics[scale=0.45]{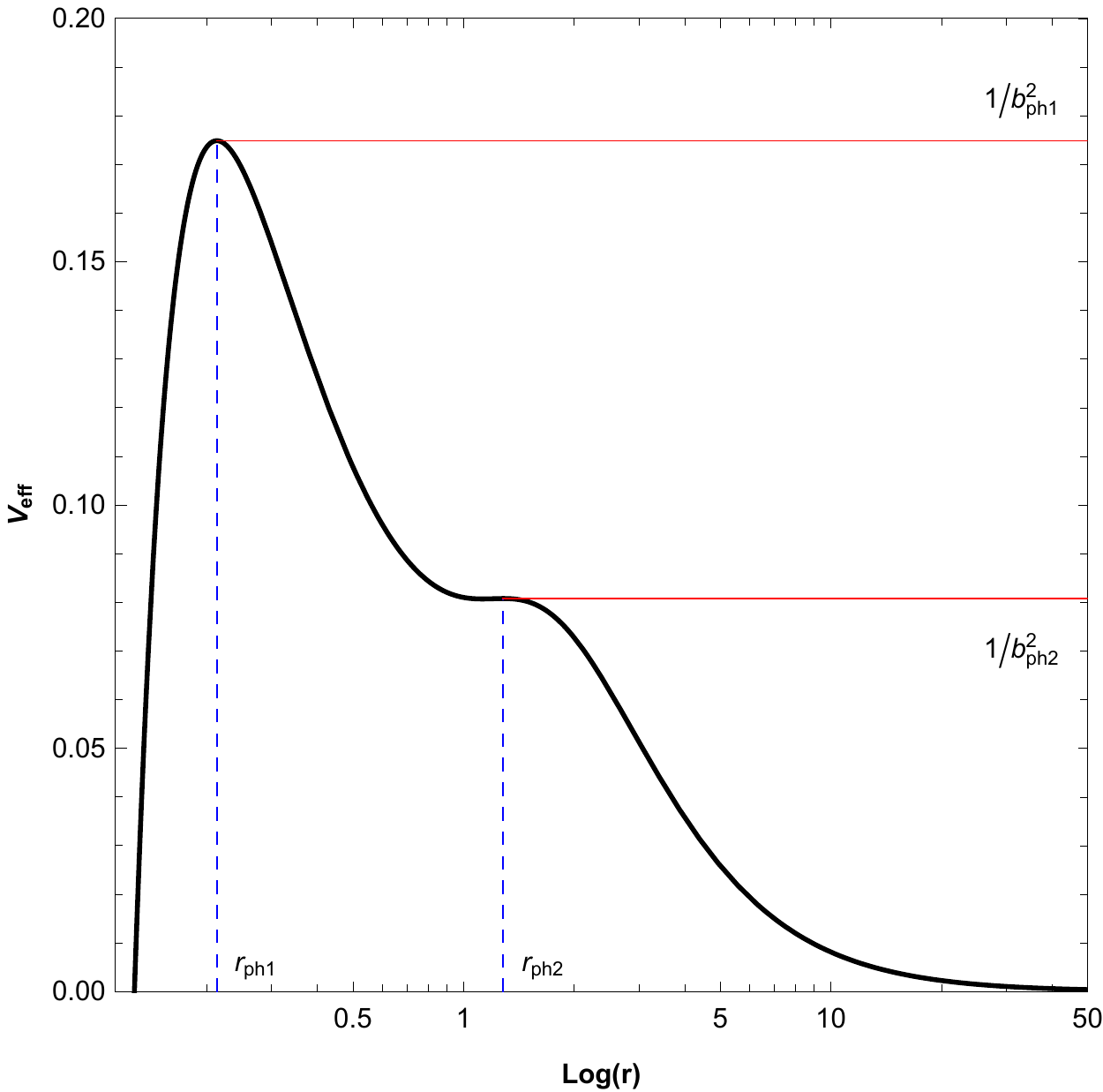}$\,$\includegraphics[scale=0.6]{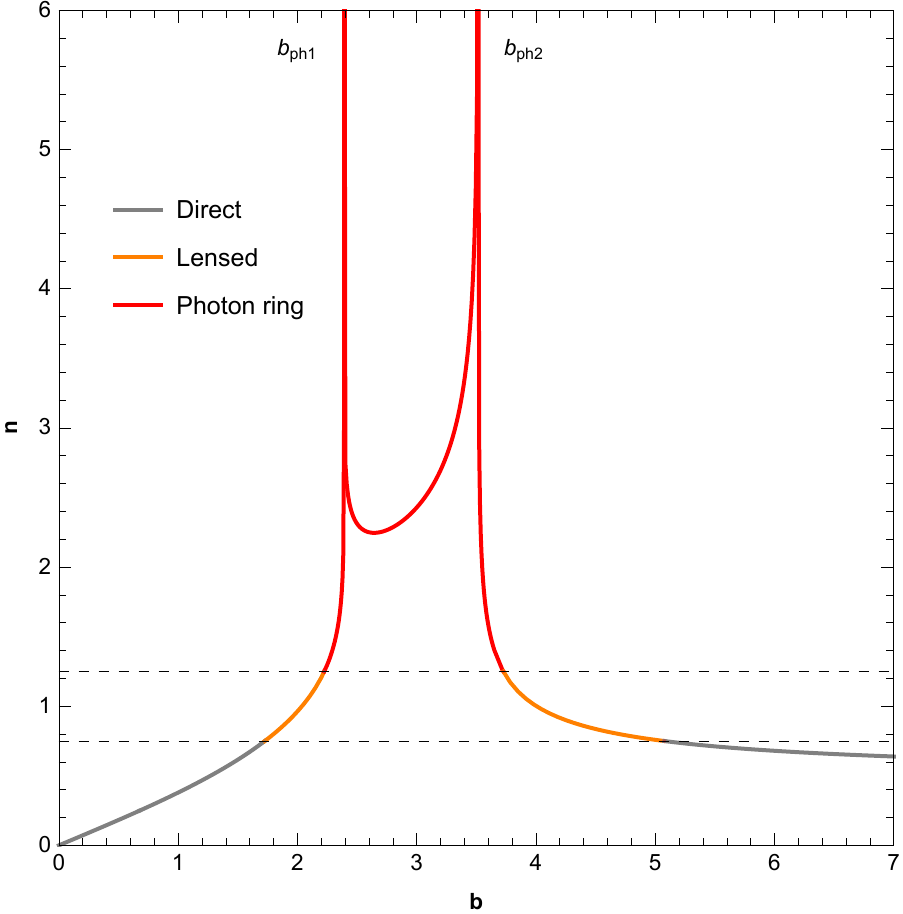}$\,$\includegraphics[scale=0.61]{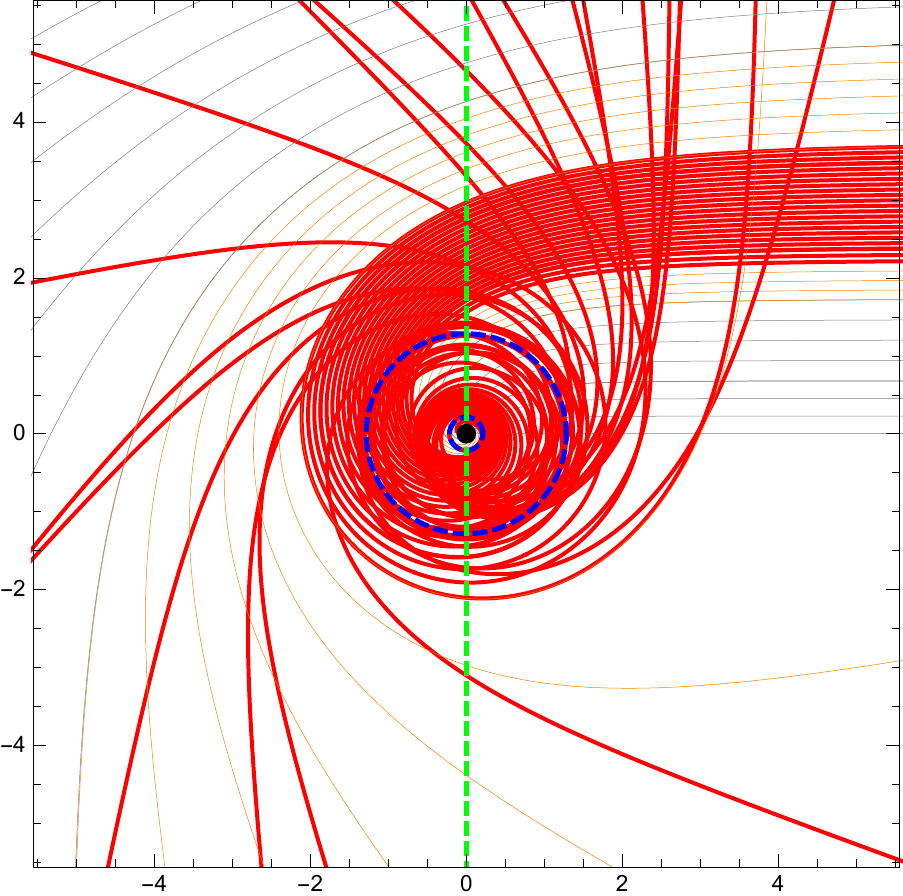}
\par\end{centering}
\vspace{5mm} \begin{centering}
\includegraphics[scale=0.42]{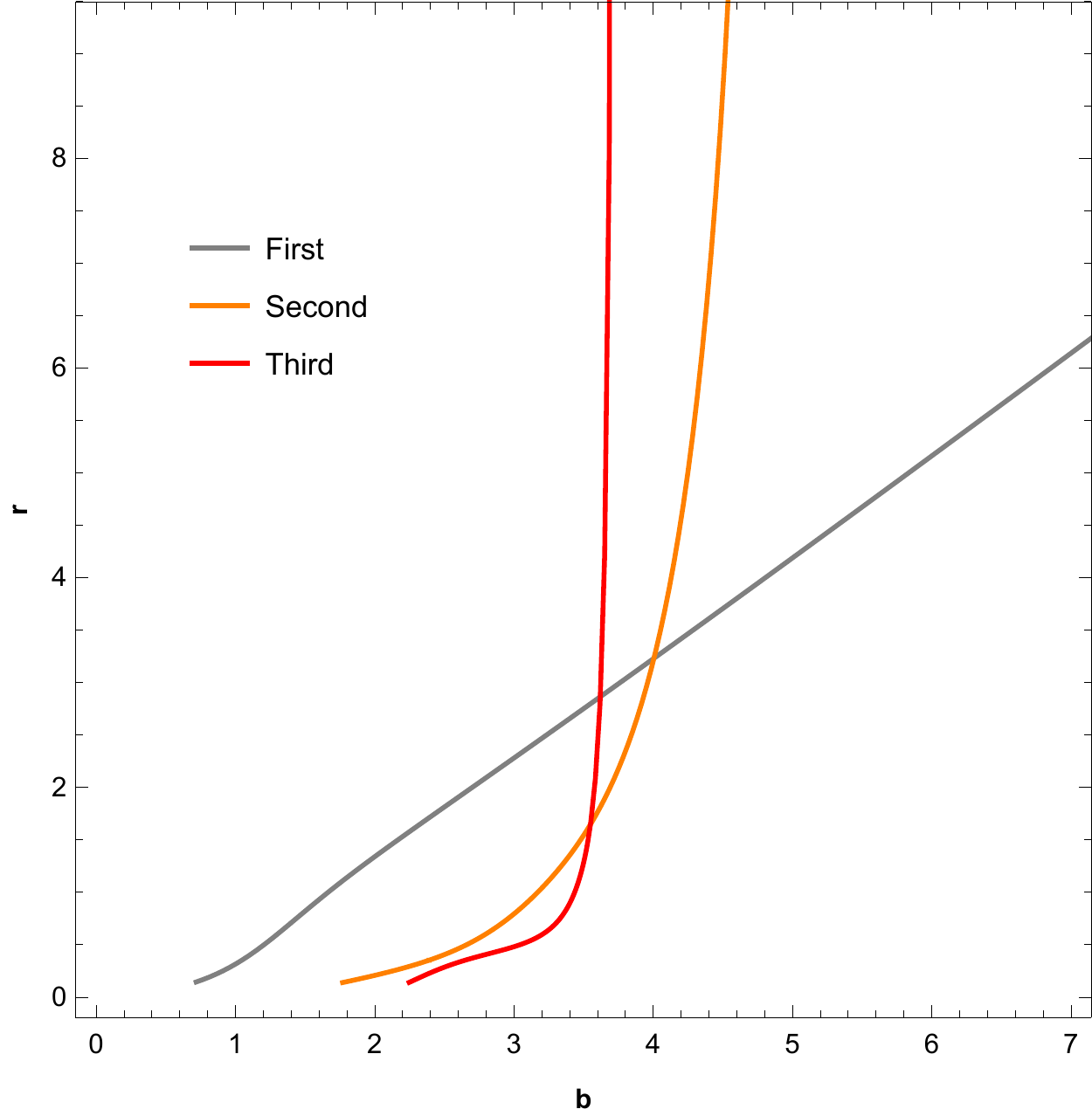}$\,$\includegraphics[scale=0.43]{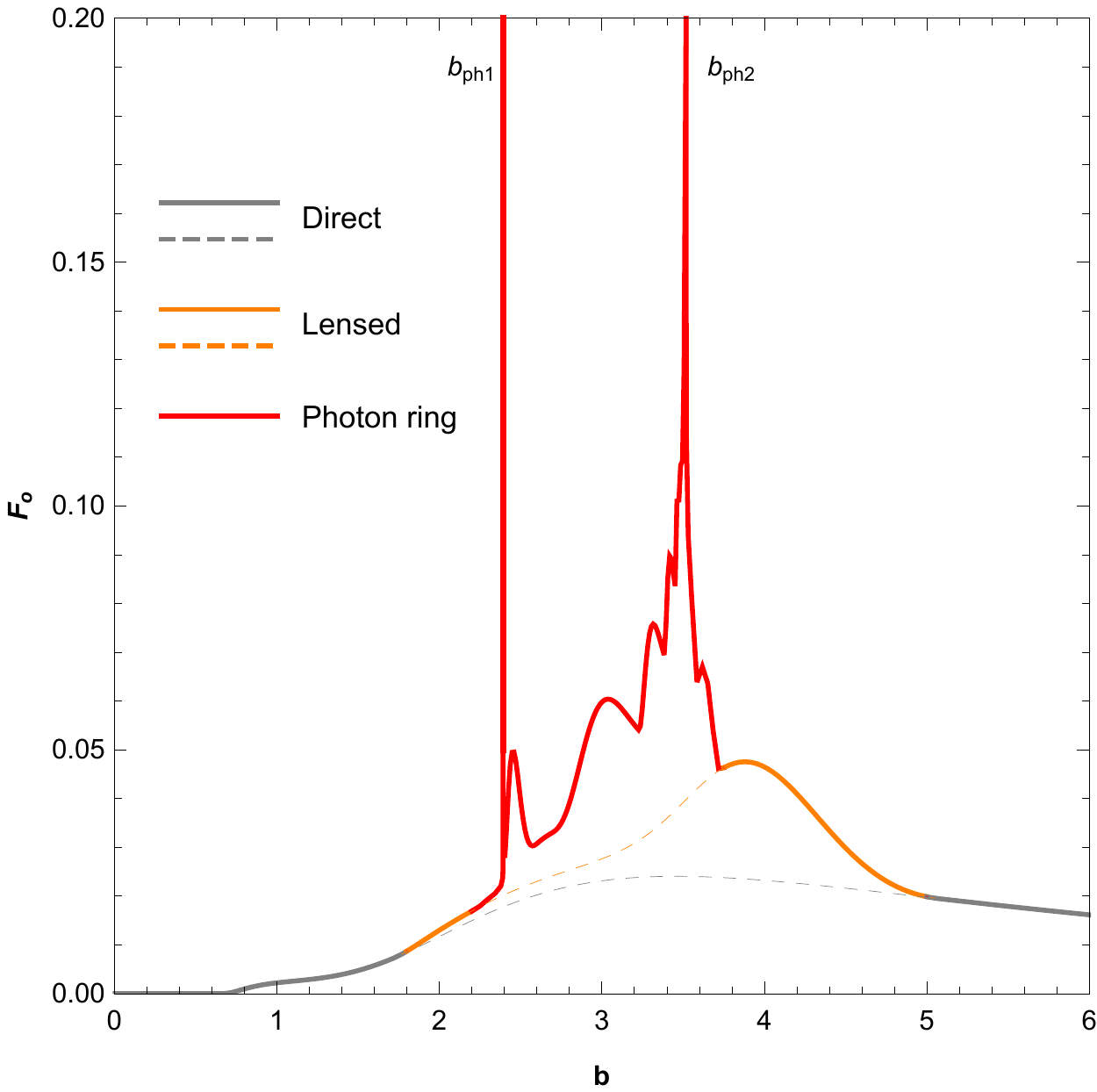}$\,$\includegraphics[scale=0.585]{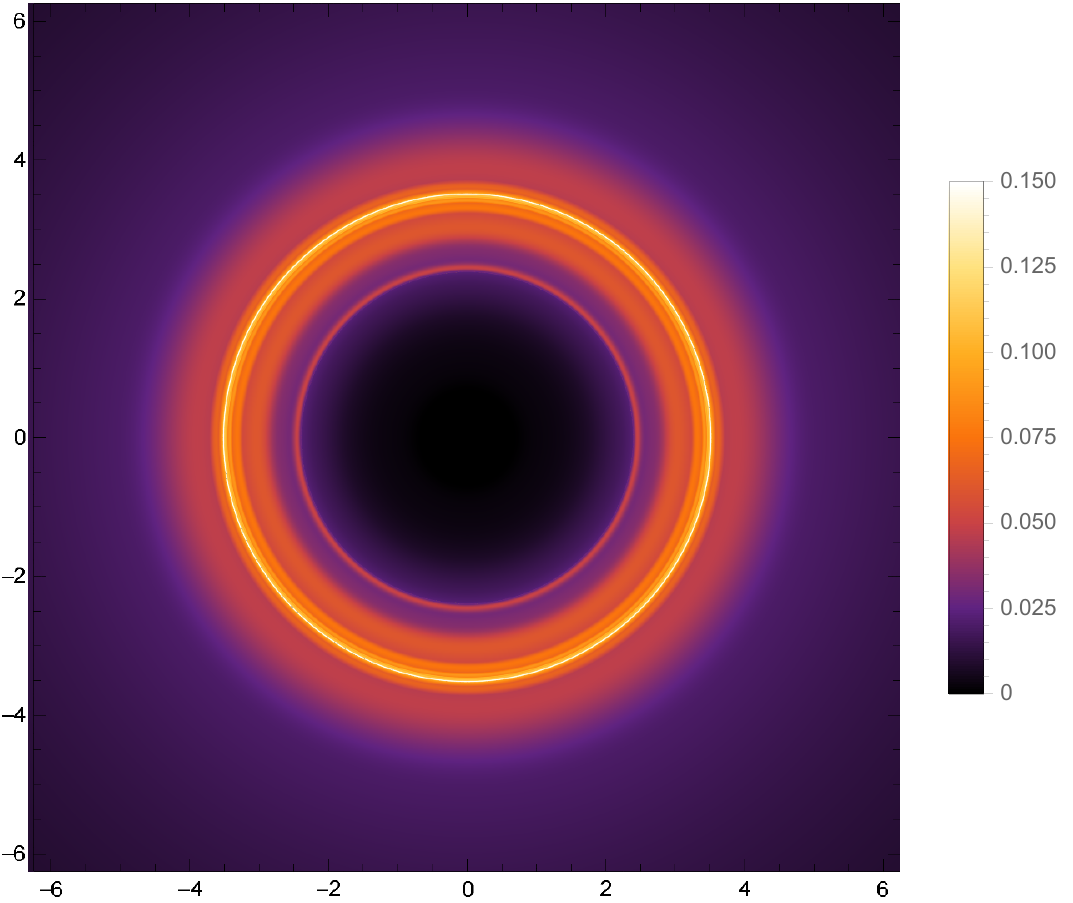}
\par\end{centering}
\caption{Behavior of photons in a HBH with $\alpha=0.9$ and $Q=1.07$, which is
in the double-peak II family, and the observational appearance of an optically
and geometrically thin accretion disk around the HBH, viewed from a face-on
orientation. Compared to Fig. \ref{figure double peak 1}, the photon ring here
becomes significantly wider (\textbf{Upper Middle and Lower Middle}) since the
two maximum values of the effective potential are well separated
(\textbf{Upper Left}). Therefore, the photon ring can make a sizable
contribution to the total flux (\textbf{Lower Middle}), and the internal
structure of the photon ring, which is made up of multiple concentric bright
rings, can be observed in the 2D image (\textbf{Lower Right}). }%
\label{figure double peak 2}%
\end{figure}

The prominent character of a double-peak potential is the presence of two
maxima at $r=r_{ph1}$ and $r=r_{ph2}$ with $r_{ph1}<r_{ph2}$, corresponding to
two photon spheres of radii $r_{ph1}$ and $r_{ph2}$ outside the horizon,
respectively. The associated maximum values of the potential are
$1/b_{ph1}^{2}$ and $1/b_{ph2}^{2}$, where $b_{ph1}$ and $b_{ph2}$ are the
impact parameters of the photon spheres (see the upper-left panels of Fig.
\ref{figure double peak 1} and \ref{figure double peak 2}). As mentioned
before, there are two families of double-peak potentials according to the
magnitudes of the two maximum values of the potentials. For the double-peak I
potential, the maximum at the smaller radius is lower than that at the larger
radius (or equivalently, $b_{ph1}>b_{ph2}$). So the light rays revolving
around the photon sphere of smaller radius can not escape to the infinity,
rendering this photon sphere invisible to a distant observer. As a result,
accretion disk images in the double-peak I family closely resemble those in
the single-peak I family. On the other hand, both photon spheres can be
responsible for obtaining the image of an accretion disk in the double-peak II
family. In fact, two bight rings of arbitrarily large brightness, whose radii
are $b_{ph1}$ and $b_{ph2}$, respectively, can appear in the image of the
accretion disk. In what follows, we display two representative cases of the
double-peak II family.

In Fig. \ref{figure double peak 1}, we consider a HBH with $\alpha=0.9$ and
$Q=1.064$, for which the difference between $b_{ph1}$ and $b_{ph2}$ is small.
As expected, the existence of two photon spheres endows both $n(b)$ and
$F_{o}(b)$ curves with two infinite peaks at radii $b_{ph1}$ and $b_{ph2}$. As
shown in the upper-middle panel, this two-peak structure extends the width of
the photon ring band, compared to the single-peak I case. It is observed in
the lower-middle panel that the photon ring can make a non-negligible
contribution to the observed intensity, which is comparable to that of the
lensing ring. Moreover, the photon ring leads to some internal structure
between the two peaks for the observed intensity. However, since the photon
ring is not wide enough, the lower-right panel shows that the internal
structure can hardly been seen in the 2D image, and the observational
appearance of the photon ring is almost indistinguishable from that of a thin
bright ring. 

In Fig. \ref{figure double peak 2}, we consider another HBH with $\alpha=0.9$
and $Q=1.07$, for which the difference between $b_{ph1}$ and $b_{ph2}$ is
large. As shown in the upper-left panel, the two maximum values of the
effective potential are quite different, thus leading to a remarkably wider
photon ring. The lower-middle panel exhibits that the lensing and photon rings
both contribute appreciably to the total observed intensity, and the photon
ring is notably wider than that in Fig. \ref{figure double peak 1}. In the
lower-right panel, the 2D observed image is shown to have a bright thin ring
at $b=b_{ph1}$ and a bright annulus around $b=b_{ph2}$, which consists of a
bright ring at $b=b_{ph2}$ and multiple concentric bright rings with different
luminosity. In other words, the inner structure of the photon ring can be seen
in the observed image since the photon ring is wide enough. From the
lower-left panels in Figs. \ref{figure double peak 1} and
\ref{figure double peak 2}, it shows that the secondary and tertiary images in
Fig. \ref{figure double peak 2} are less demagnified than those in Fig.
\ref{figure double peak 1}. It is noteworthy that the ankle-like structure of
the potential in Fig. \ref{figure single peak 2} is reminiscent of the maximum
of the potential at $r=r_{ph2}$ in Fig. \ref{figure double peak 2}.

\subsection{Dependence on black hole charge and scalar coupling}

\label{subsec:Fixed alpha or Q}

\begin{figure}[ptb]
\begin{centering}
\includegraphics[scale=0.63]{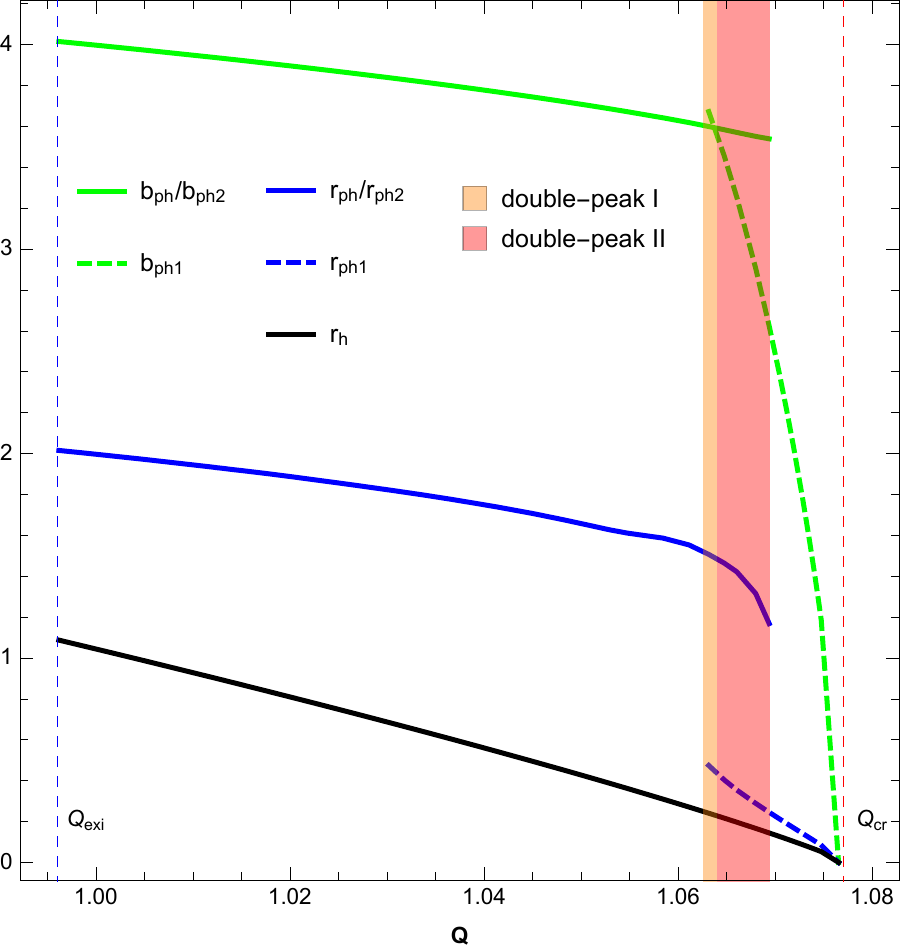}$\;$\includegraphics[scale=0.655]{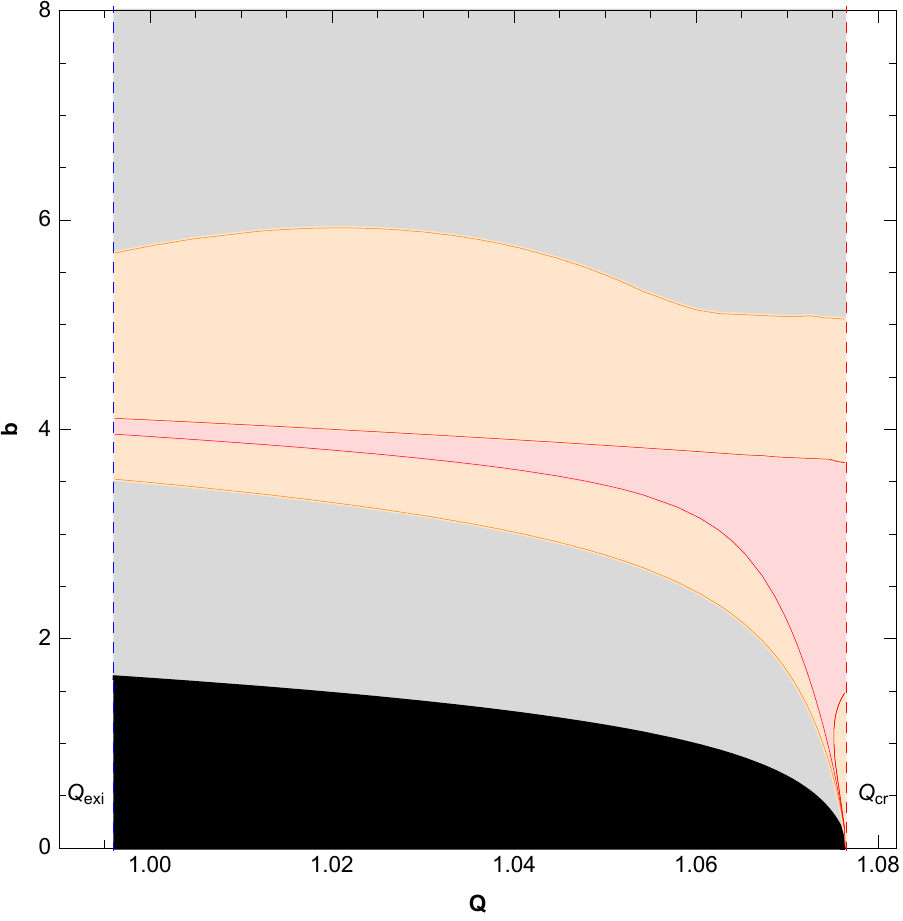}$\;$\includegraphics[scale=0.462]{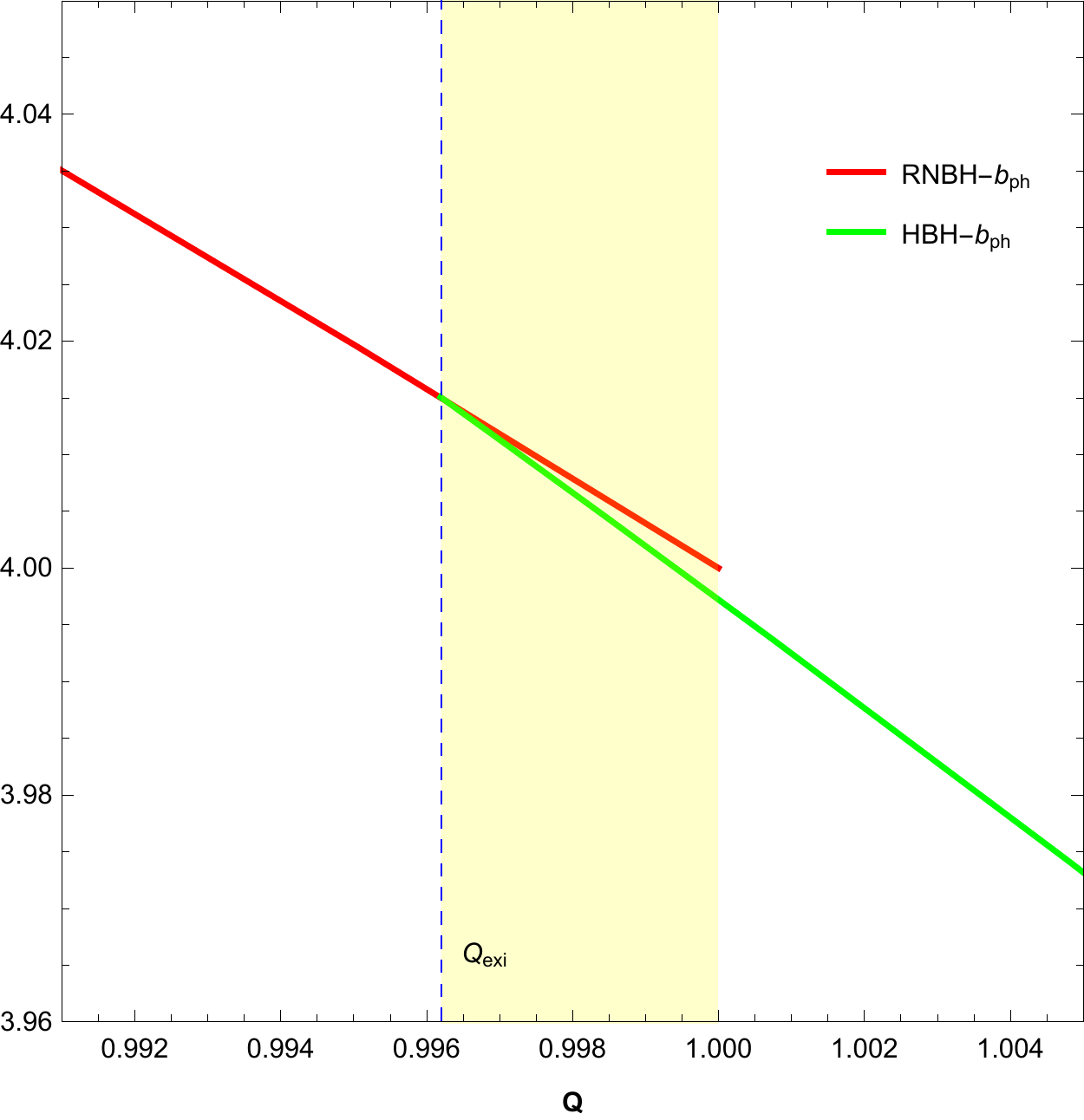}
\par\end{centering}
\vspace{5mm} \begin{centering}
\includegraphics[scale=0.45]{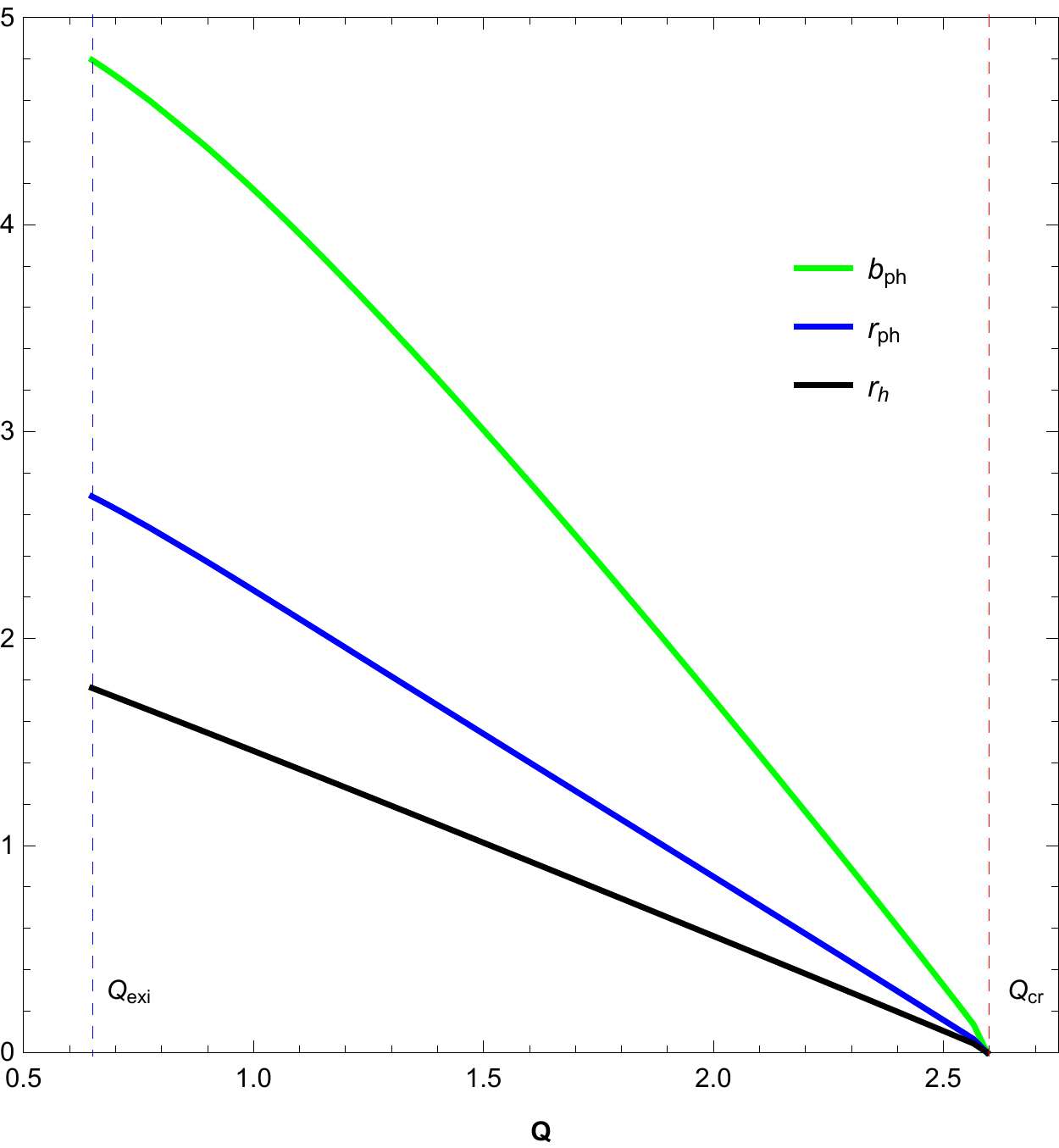}$\;$\includegraphics[scale=0.655]{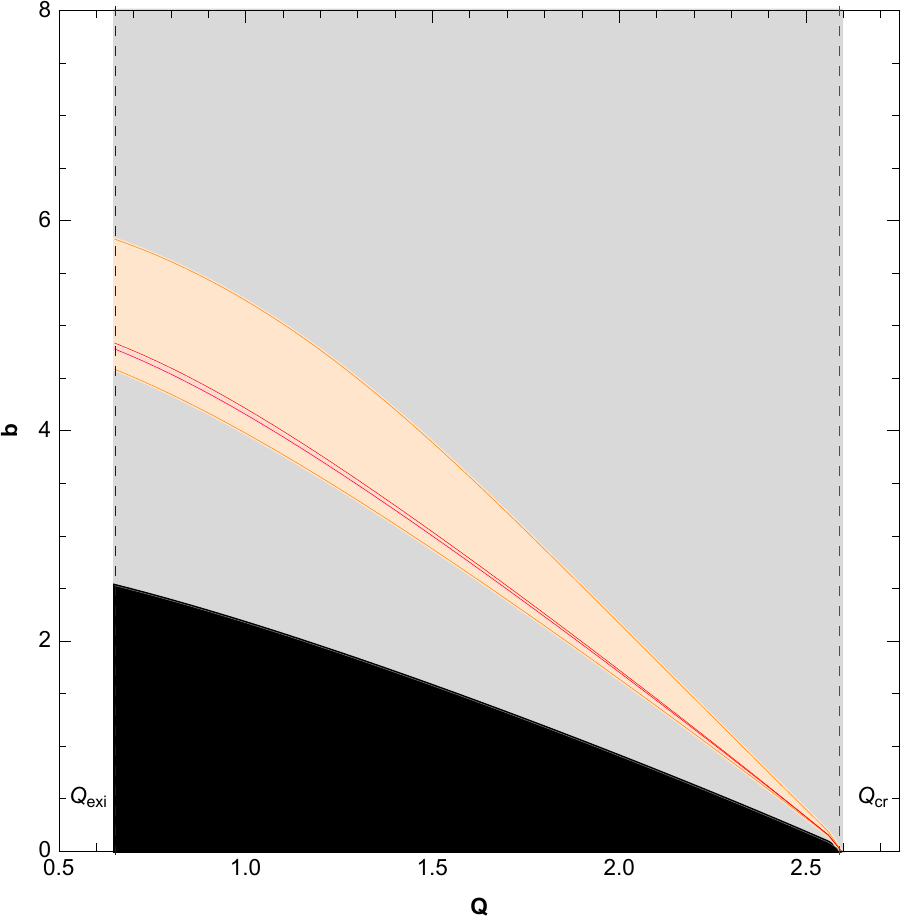}$\;$\includegraphics[scale=0.462]{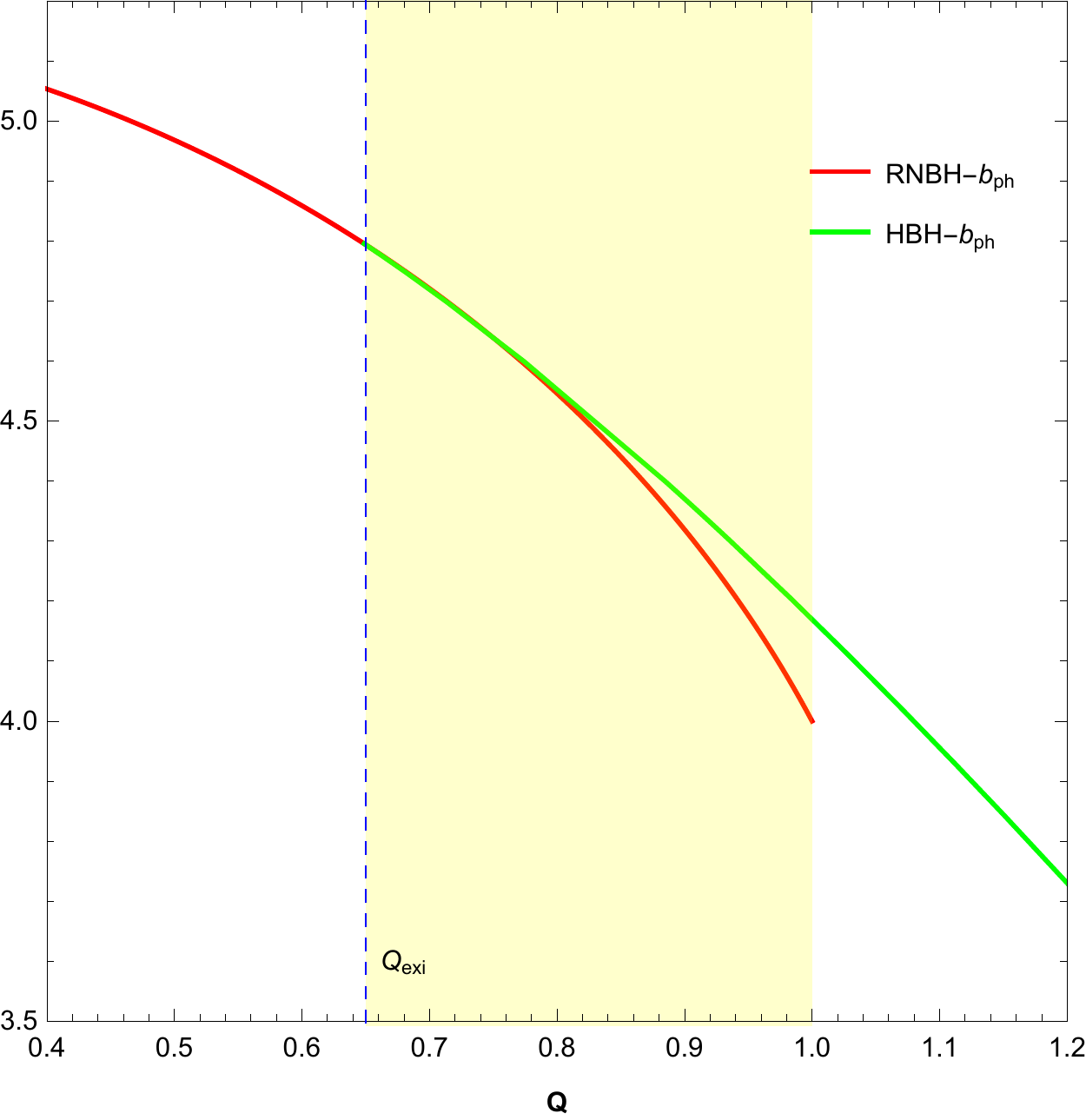}
\par\end{centering}
\caption{Dependence of quantities of interest on the black hole charge $Q$ for
HBHs with $\alpha=0.9$ ($\mathbf{Upper}$) and $\alpha=10$ ($\mathbf{Lower}$).
$\mathbf{Left}$: The photon sphere radii, $r_{ph}$, $r_{ph1}$ and $r_{ph2}$,
the associated impact parameters, $b_{ph}$, $b_{ph1}$ and $b_{ph1}$, and the
horizon radius $r_{h}$ as functions of $Q$. As shown in Fig. \ref{figure veff}%
, the range of $Q$ is bounded by the existence charge $Q_{\text{exi}}$ (black
dashed) and critical charge $Q_{\text{cr}}$ (red dashed). With increasing $Q$
from $Q_{\text{exi}}$, the HBH with $\alpha=0.9$ undergoes the single-peak I
(region between the $Q_{\text{exi}}$-line and the orange region), double-peak
I (orange region), double-peak II (red region) and the single-peak II (region
between the red region and the $Q_{\text{cr}}$-line) families, while the HBH
with $\alpha=10$ remains in the single-peak I family. The impact parameter of
the (smaller) photon sphere is identified as the radius of the HBH standard
shadow, which decreases in size as $Q$ increases, and disappears at
$Q=Q_{\text{cr}}$. $\mathbf{Middle}$: The ranges of the direct emission
(black, gray, orange and red regions), the lensing ring (orange and red
regions) and the photon ring (red region) as functions of $Q$. The black
region denotes the $m=0$ band, corresponding to the completely dark region.
When $\alpha=0.9$, the range of the photon ring can be quite noticeable for
large enough $Q$ (e.g., in the single-peak II and double-peak II families),
thus indicating that the photon ring contributes non-trivially to the observed
flux. When $\alpha=10$, the photon ring always has a very narrow range, and
hence makes a negligible contribution to the flux. $\mathbf{Right}$: The
impact parameters of the RNBH (red) and HBH (green) versus $Q$ in the
coexistence region (yellow region). For a given $Q$, the HBH/RNBH has a
smaller standard shadow for $\alpha=0.9/\alpha=10$. }%
\label{figure alpha fixed}%
\end{figure}

\begin{figure}[ptb]
\begin{centering}
\includegraphics[scale=0.45]{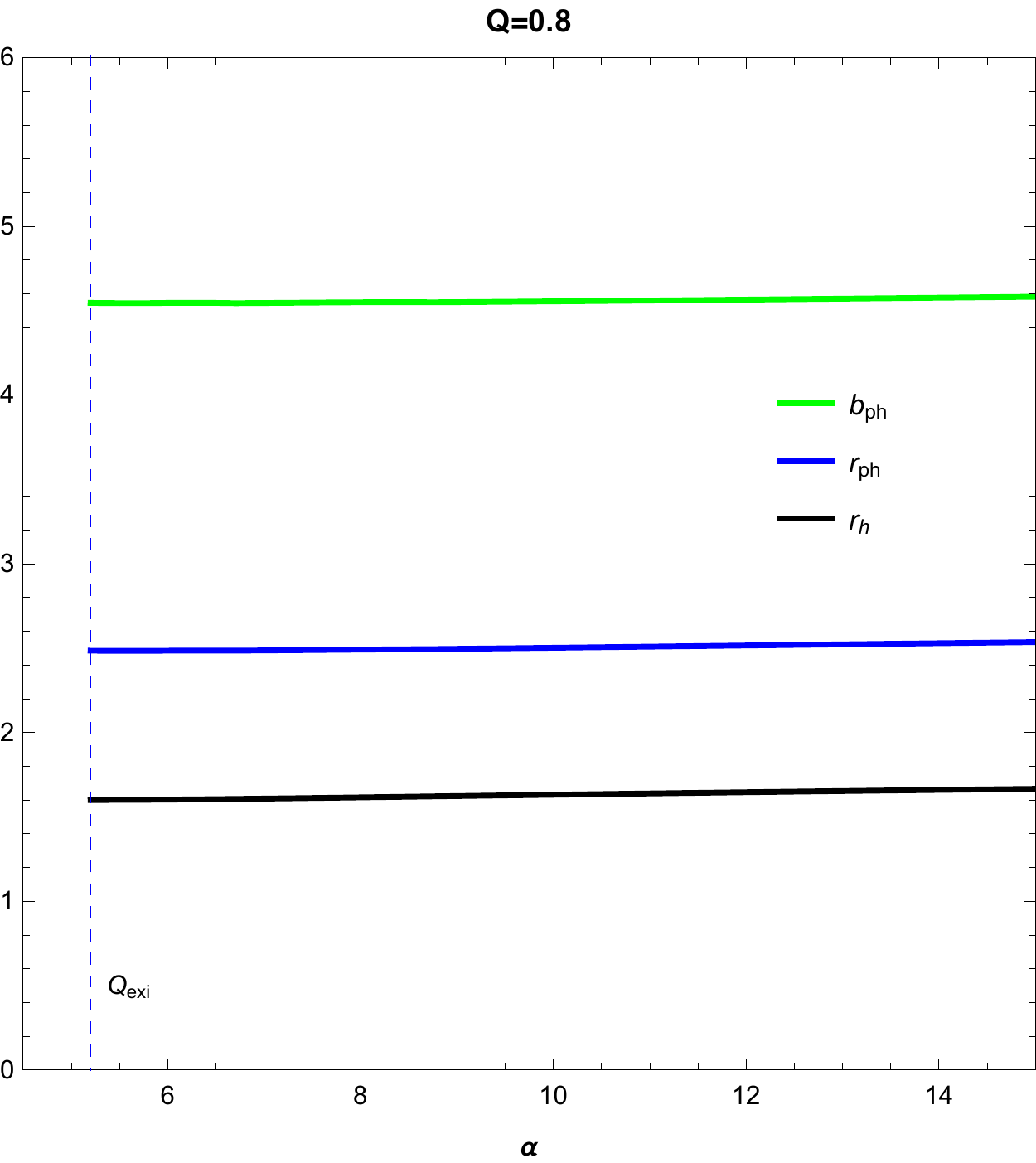}$\;$\includegraphics[scale=0.63]{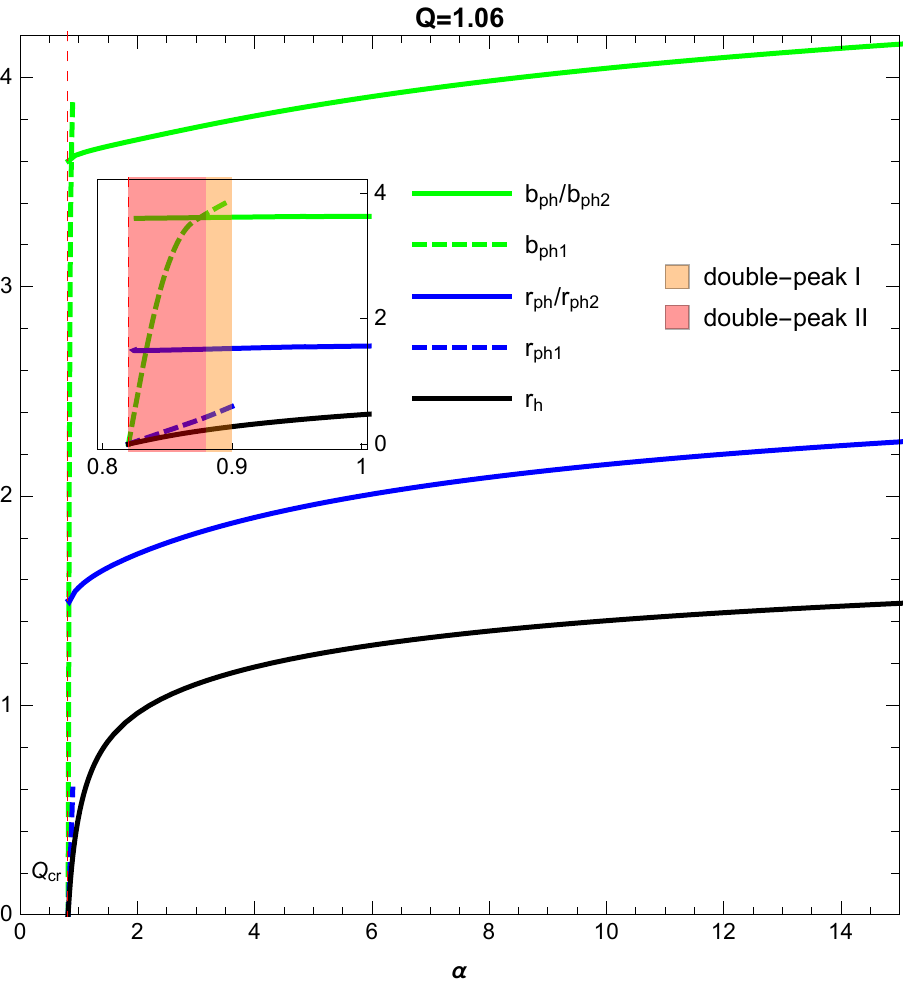}$\;$\includegraphics[scale=0.46]{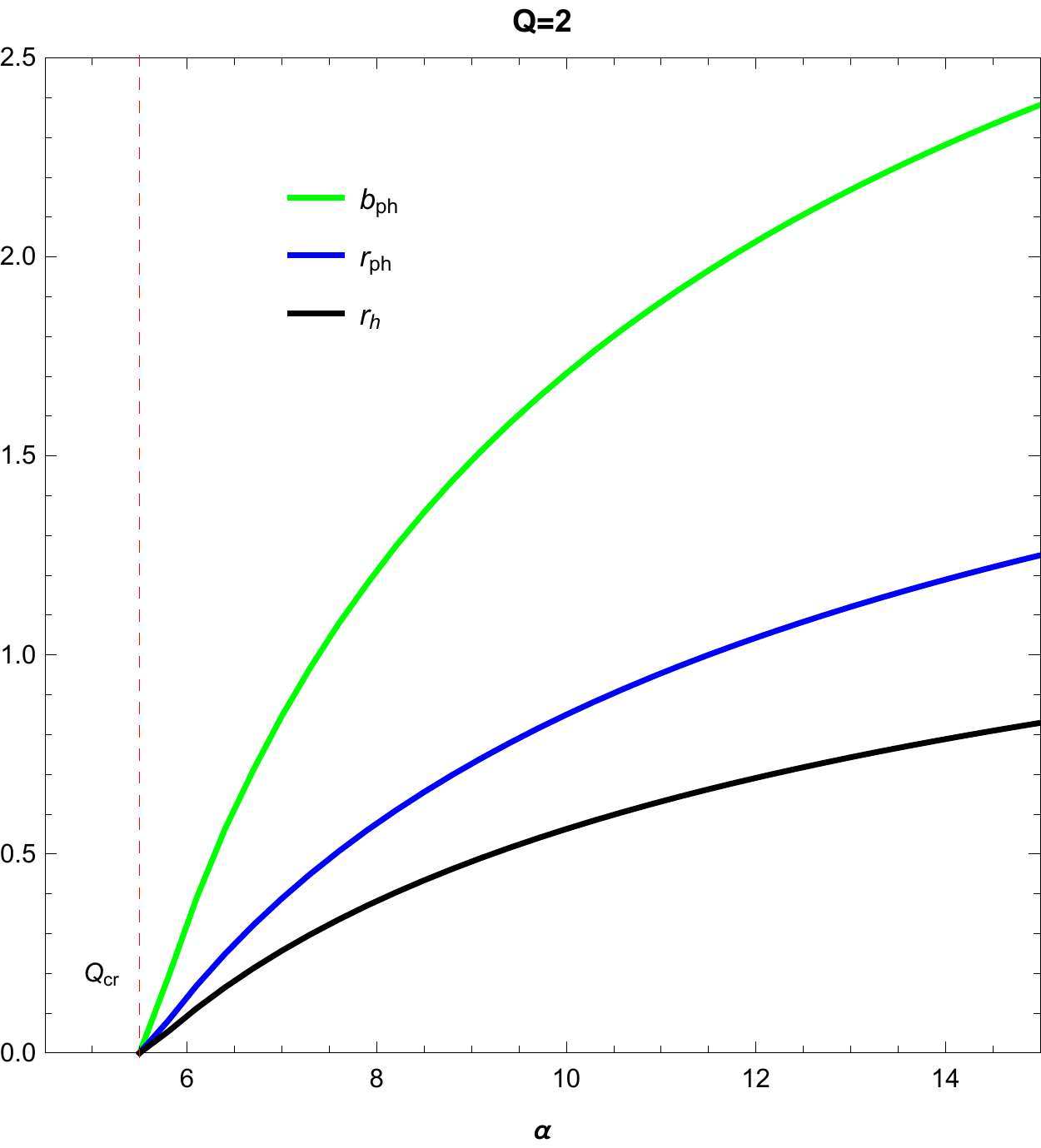}
\par\end{centering}
\vspace{5mm} \begin{centering}
\includegraphics[scale=0.625]{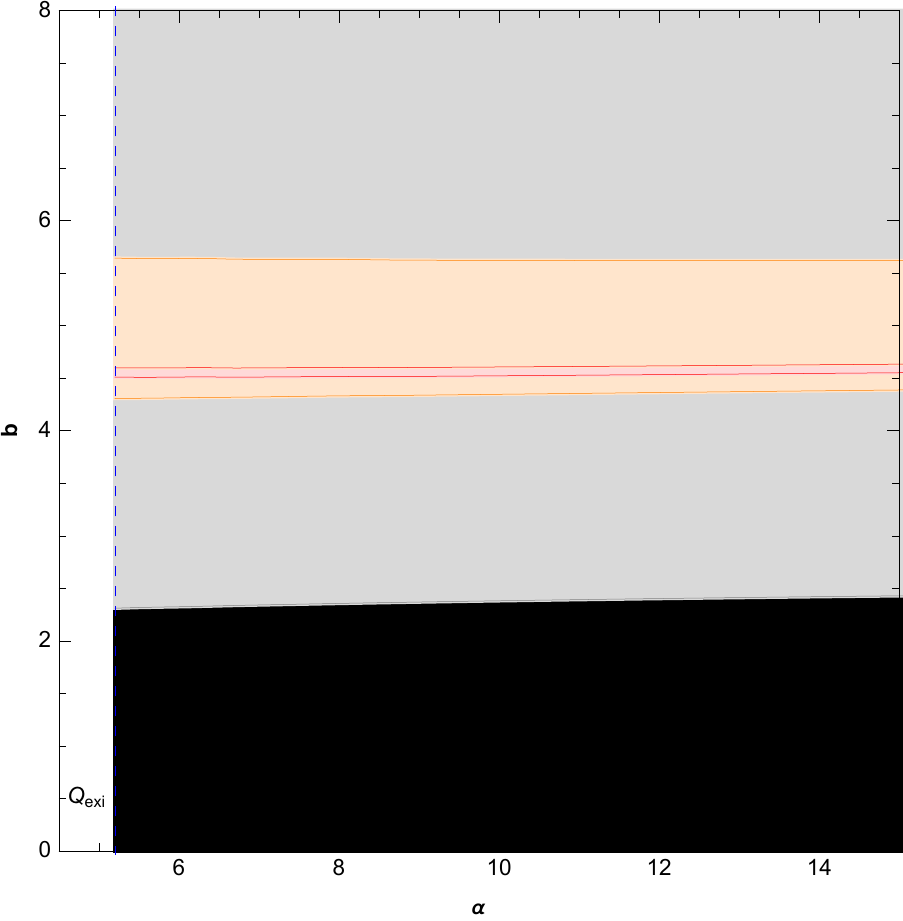}$\;$\includegraphics[scale=0.625]{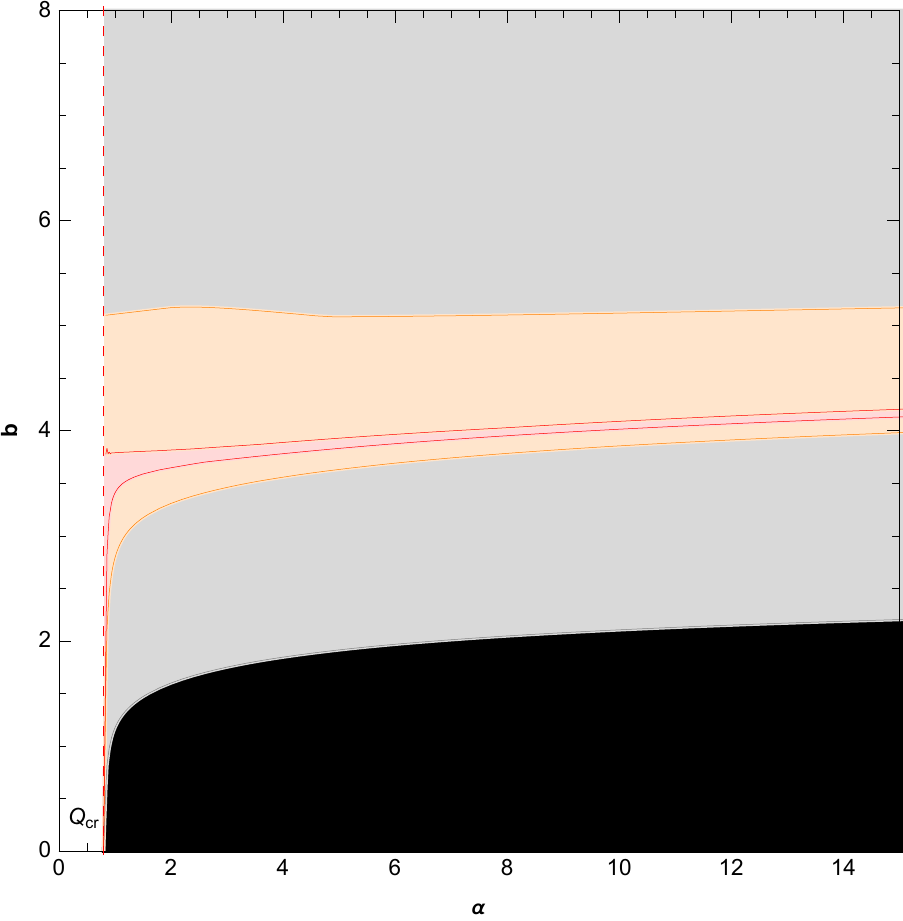}$\;$\includegraphics[scale=0.625]{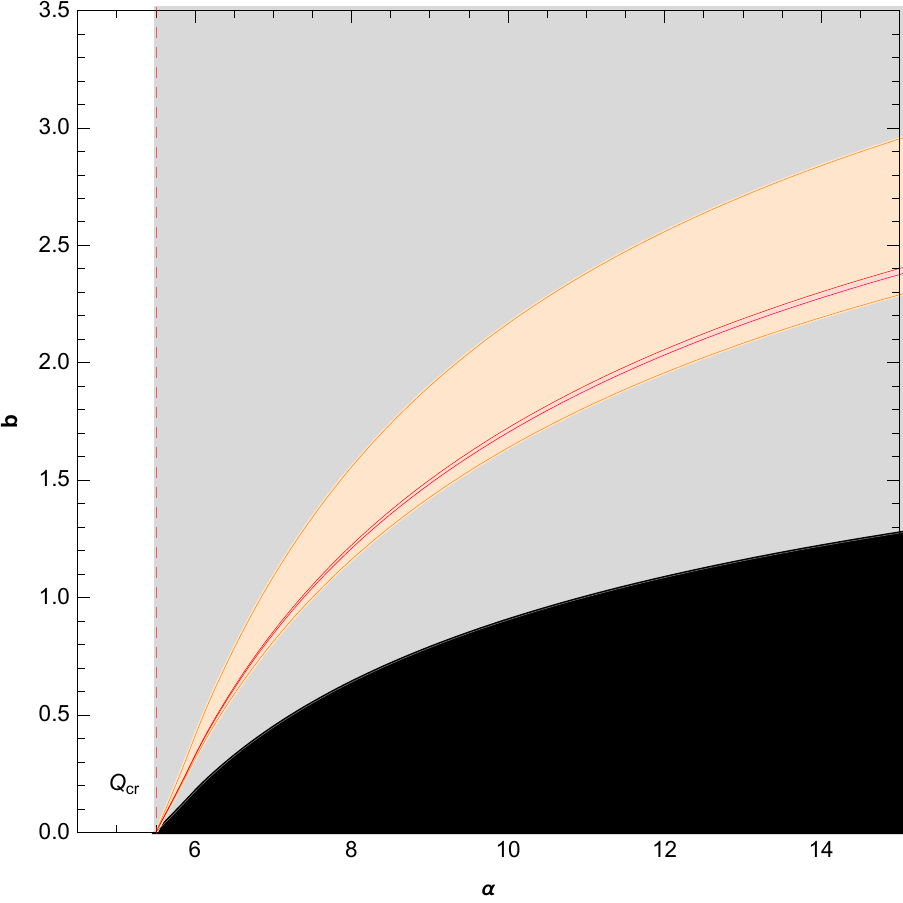}
\par\end{centering}
\caption{Dependence of quantities of interest on the scalar coupling $\alpha$
for HBHs with $Q=0.8$ ($\mathbf{Left}$), $Q=1.06$ ($\mathbf{Middle}$) and
$Q=2$ ($\mathbf{Right}$). For the $Q=0.8$ and $Q=2$ cases, the HBHs are always
in the single-peak I family. As $\alpha$ increases, the HBH with $Q=1.06$
belongs to the double-peak II, double-peak I and single-peak I families,
respectively. The allowed range of $\alpha$ is bounded from below by the
$Q_{\text{exi}}$-line in the $Q=0.8$ case, and by the $Q_{\text{cr}}$-line in
the $Q=1.06$ and $Q=2$ cases. $\mathbf{Upper}$: The event horizons radii
(black), the photon spheres radii (blue) and the corresponding impact
parameters (green) versus $\alpha$. The standard shadow grows as $\alpha$
increases. $\mathbf{Lower}$: The ranges of the direct emission (black, gray,
orange and red regions), the lensing ring (orange and red regions) and the
photon ring (red region) as functions of $\alpha$. The black region denotes
the completely dark region. The photon ring can play a relevant role for a HBH
with $Q=1.06$, that is close enough to the $Q_{\text{cr}}$-line.}%
\label{figure Q fixed}%
\end{figure}

Here, we turn to investigate the dependence of the size of the photon ring and
shadow on the HBH charge $Q$ and the scalar coupling $\alpha$. To study the
dependence on $Q$, we consider two cases with fixed $\alpha=0.9$ and
$\alpha=10$ in the upper and lower rows of Fig. \ref{figure alpha fixed},
respectively. Note that the allowed parameter regions are bounded by the
$Q_{\text{exi}}$-line and/or the $Q_{\text{cr}}$-line, which are shown in Fig.
\ref{figure veff}. The upper-left panel of Fig. \ref{figure alpha fixed}
displays that, when $\alpha=0.9$, the HBHs belong to the single-peak I (region
on the left of the orange region), single-peak II (region on the right of the
red region), double-peak I (orange region) or double-peak II (red region)
families, depending on the value of the HBH charge. From the upper middle
panel, one can see that the photon ring (red region) of the HBHs in the
single-peak II and double-peak II families are quite wide, thus indicating
that the photon rings can play a relevant role in determining the observed
accretion disk images. Moreover, as the HBH charge increases toward the
$Q_{\text{cr}}$-line, the width of the photon ring grows until the photon ring
splits into a narrow photon ring of smaller radius and a broad one of larger
radius in the image plane. While the photon ring of smaller radius is barely
visible in the observed 2D image, the one of larger radius can make a
non-negligible contribution to the total intensity and the accretion disk
image. Therefore, it is expected that the accretion disk image in the two
photon rings scenario is quite similar to those shown in Figs.
\ref{figure single peak 2} and \ref{figure double peak 2}. On the other hand,
the HBH with $\alpha=10$ is always in the single-peak I family, which is shown
in the inset of the left panel of Fig. \ref{figure veff}. In this case, the
lower-middle panel of Fig. \ref{figure alpha fixed} displays that the photon
ring is always very narrow, and the lensing (orange and red regions) and
photon rings decrease in size with $Q$ increasing toward $Q_{\text{cr}}$.
Around $Q=Q_{\text{cr}}$, the lensing ring also becomes very narrow, and makes
a negligible contribution to the observed intensity.

In this paper, the term \textquotedblleft standard shadow\textquotedblright%
\ is used to refer to the area inside the (smaller) photon sphere (i.e., the
apparent boundary \cite{Bardeen:1972fi,Bardeen:1973tla}). Specifically, the
radius of the standard shadow is the impact parameter of the smaller photon
sphere if there exist two photon spheres \cite{Gan:2021pwu}. From the left
column of Fig. \ref{figure alpha fixed}, it is observed that the standard
shadow of a HBH in the $\alpha=0.9$ and $\alpha=10$ cases shrinks with
increasing $Q$ and vanishes at $Q=Q_{\text{cr}}$, where the HBH horizon
becomes zero. For $\alpha=0.9$, the standard shadow radius decreases at a
larger decreasing rate in the single-peak II and double-peak II families. In
the middle column of Fig. \ref{figure alpha fixed}, the black regions denote
the completely dark area with vanishing intensity, and are also shown to
decrease in size with increasing $Q$ and vanishes at $Q=Q_{\text{cr}}$. For
comparison, we also plot the impact parameters of the photon spheres of RNBHs
with $M=1$ in the right column of Fig. \ref{figure alpha fixed}. In the
coexisting $Q$-range of the RNBH and HBH, one can see that the standard shadow
of the RNBH is larger than that of the HBH in the $\alpha=0.9$ case
(upper-right panel of Fig. \ref{figure alpha fixed}), and vice verse in the
$\alpha=10$ case (lower-right panel of Fig. \ref{figure alpha fixed}), which
is consistent with the result in \cite{Konoplya:2019goy}.

In Fig. \ref{figure Q fixed}, three cases with fixed $Q=0.8$ (left column),
$Q=1.06$ (middle column) and $Q=2$ (right column) are presented to study the
dependence on the scalar coupling $\alpha$. As shown in the upper row, the
size of the standard shadow becomes larger for a stronger scalar coupling in
all cases. For $Q=0.8$, the ranges of the photon and lensing rings are fairly
insensitive to $\alpha$. In the near $Q_{\text{cr}}$-line regime, HBHs with
$Q=1.06$ are in the double-peak II family, and the photon ring's contribution
to the observed intensity can not be neglected. On the other hand, the ranges
of the photon and lensing rings of HBHs with $Q=2$ become zero at the
$Q_{\text{cr}}$-line. Finally, it is interesting to note that the stronger
$\alpha$, the larger the completely dark area becomes.

\subsection{Blurred accretion disk images}

\begin{figure}[ptb]
\begin{centering}
\includegraphics[scale=0.45]{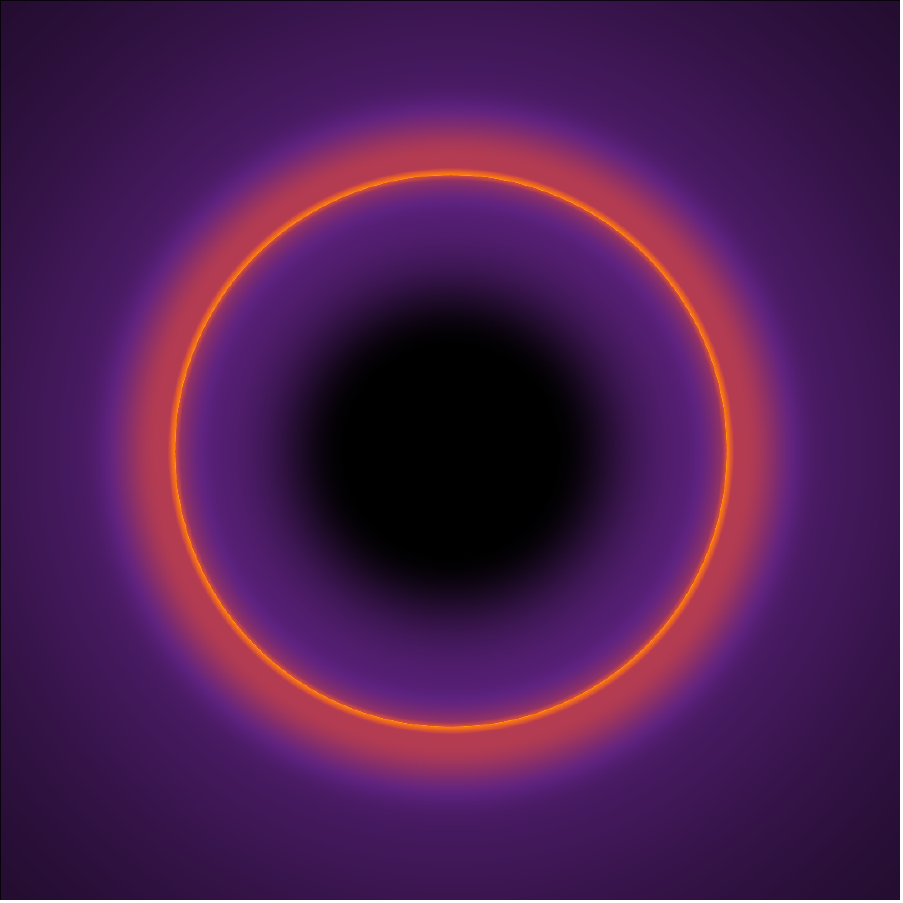}$\;$\includegraphics[scale=0.45]{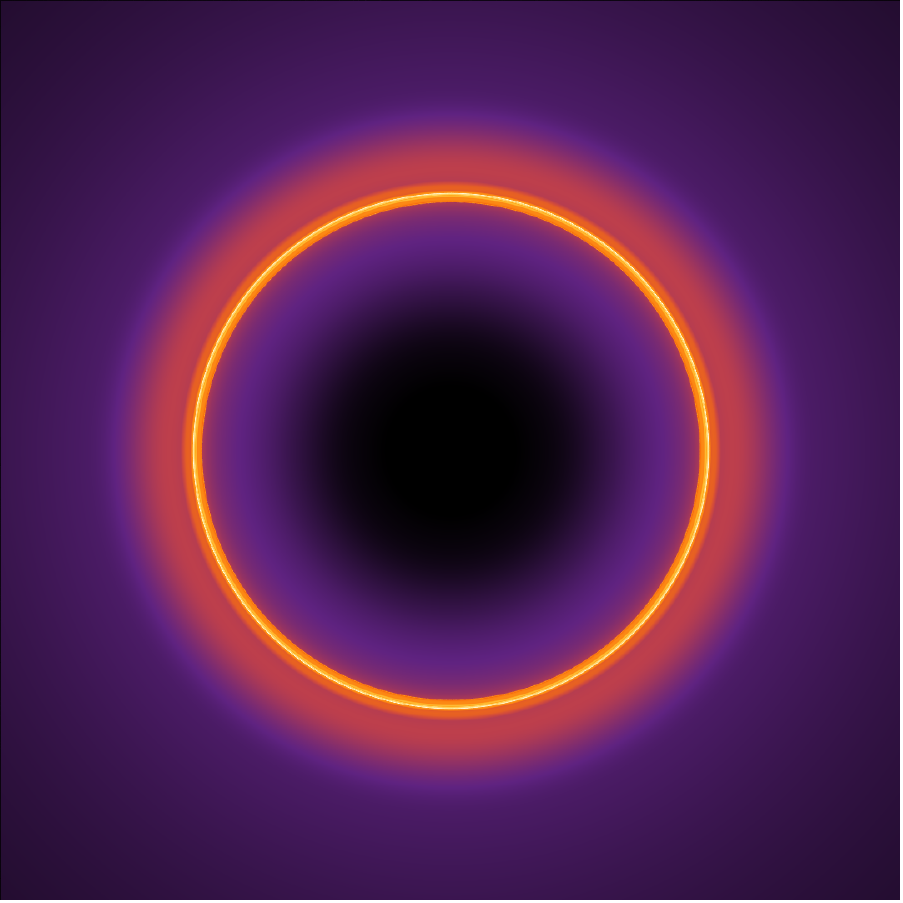}$\;$\includegraphics[scale=0.45]{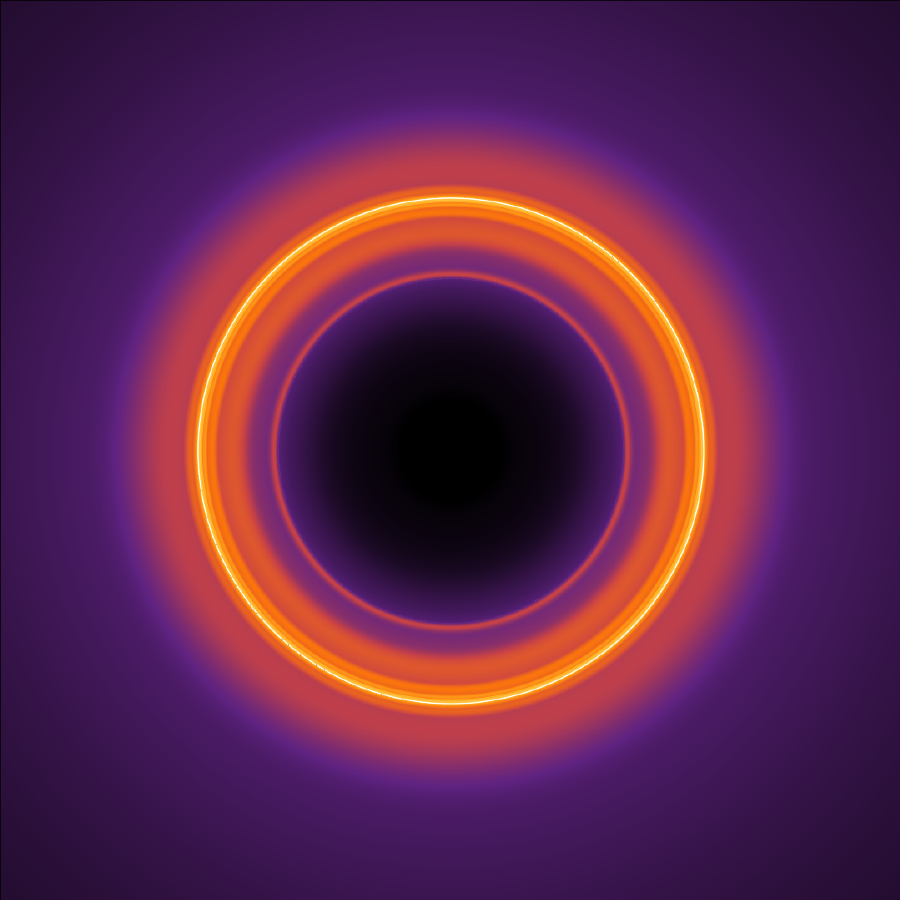}$\;$\includegraphics[scale=0.45]{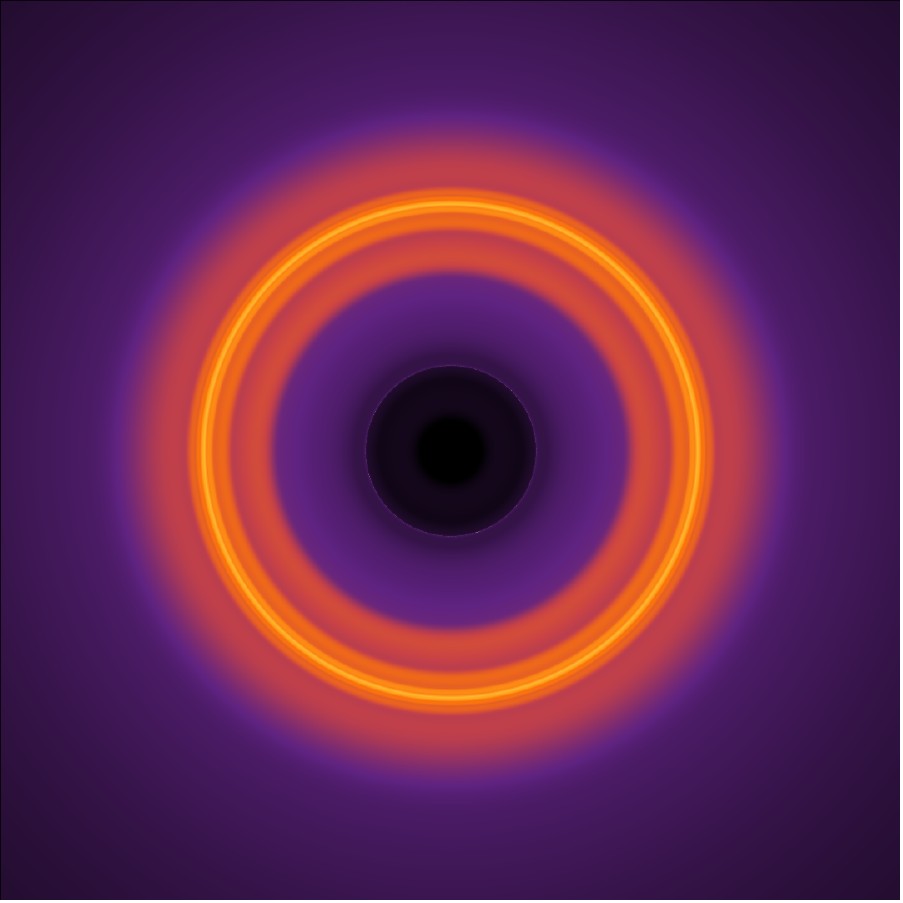}
\par\end{centering}
\begin{centering}
\includegraphics[scale=0.45]{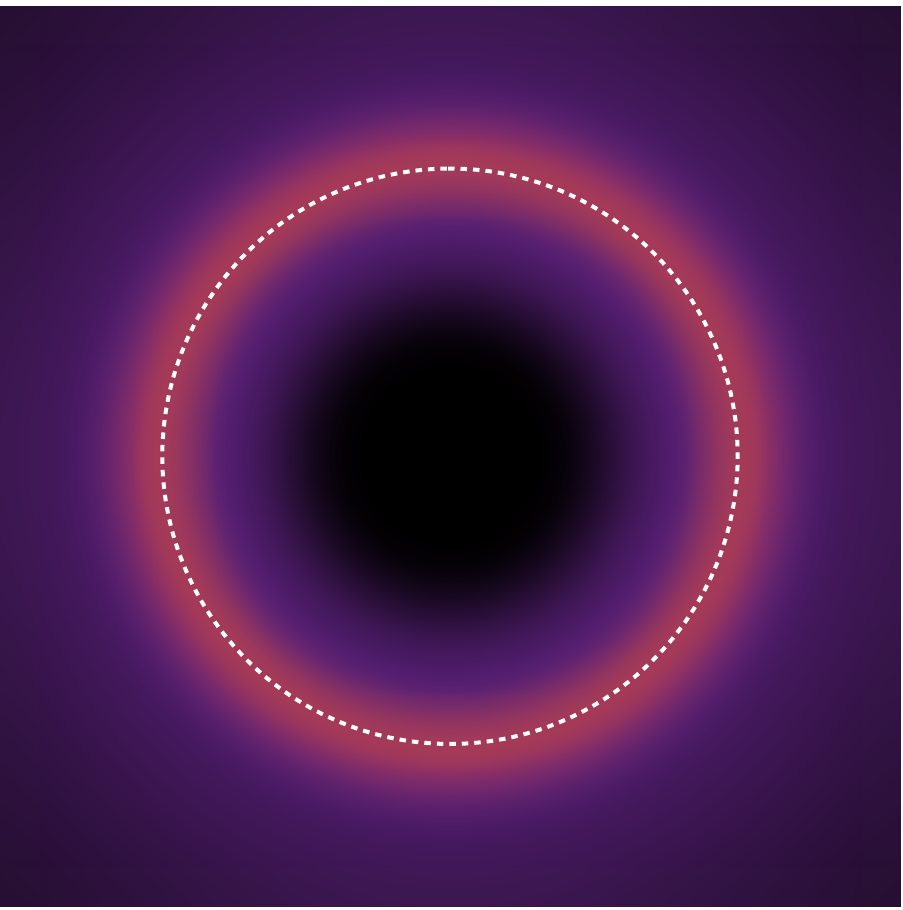}$\;$\includegraphics[scale=0.45]{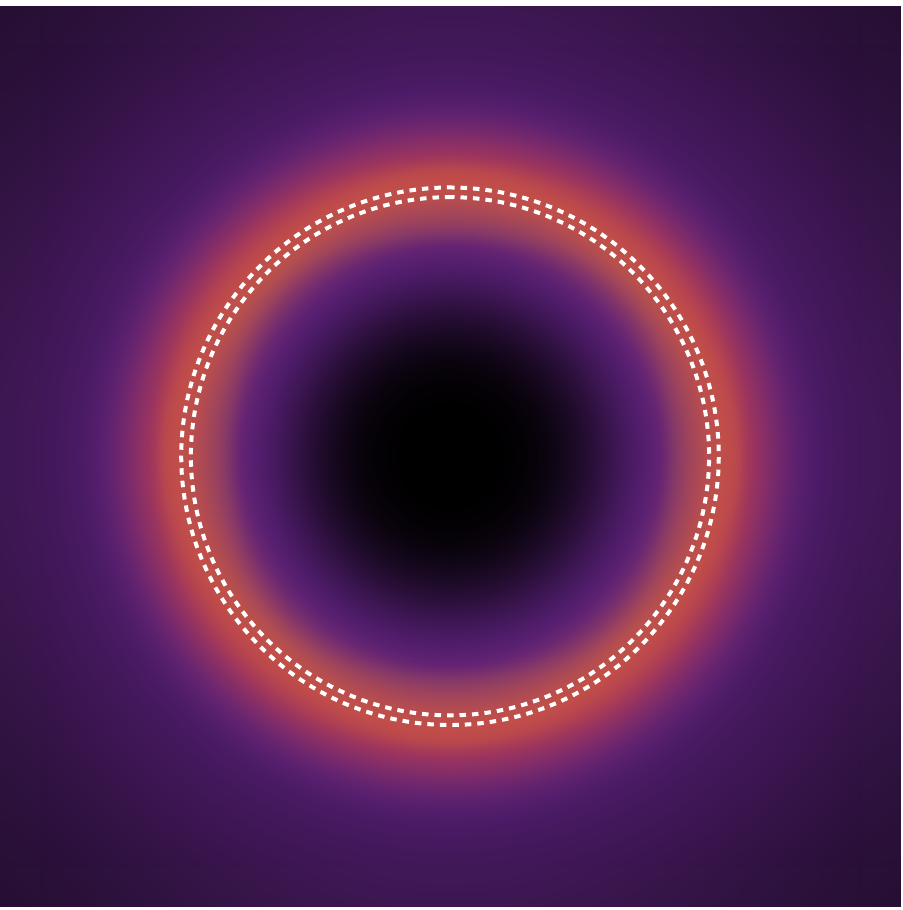}$\;$\includegraphics[scale=0.45]{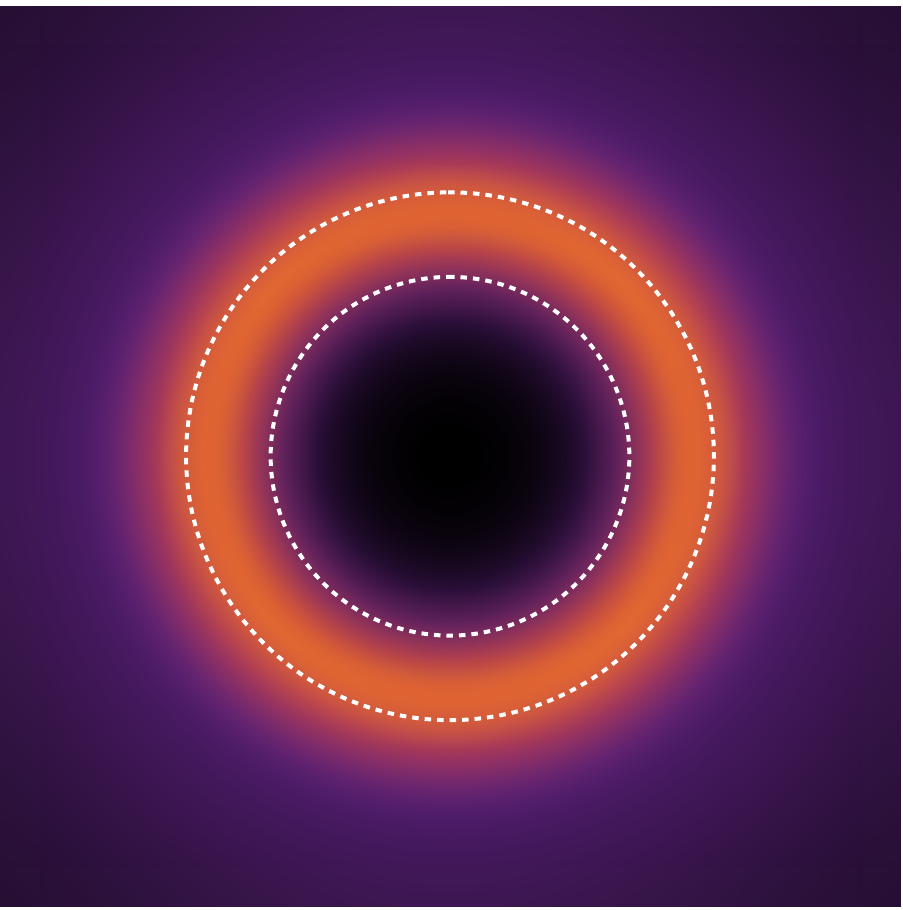}$\;$\includegraphics[scale=0.45]{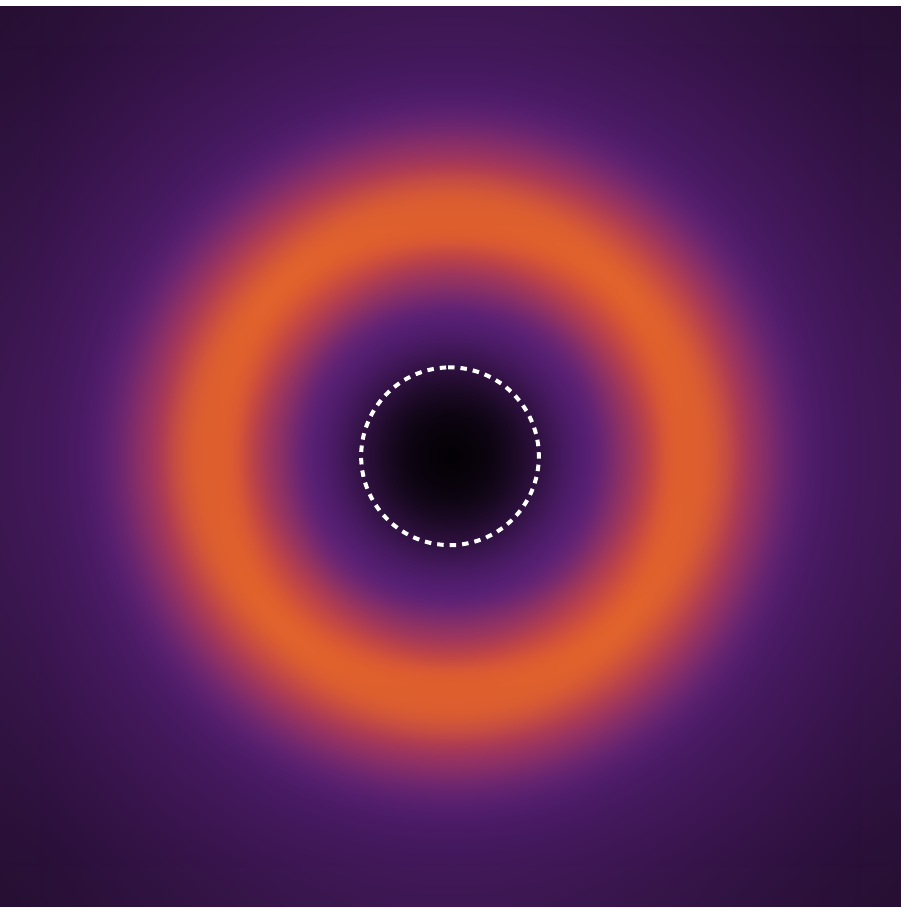}
\par\end{centering}
\caption{High resolution and blurred images of the HBHs from Fig.
\ref{figure single peak 1} (\textbf{First Column, }the single-peak I solution
with $Q=1.03$), Fig. \ref{figure double peak 1} (\textbf{Second Column, }the
double-peak II solution with $Q=1.064$), Fig. \ref{figure double peak 2}
(\textbf{Third Column, }the double-peak II solution with $Q=1.07$) and Fig.
\ref{figure single peak 2} (\textbf{Last Column, }the single-peak II solution
with $Q=1.074$). The high resolution images are blurred to correspond roughly
to the EHT resolution. The white dotted circles denote critical circles
associated with photon spheres, and the (smaller) critical circle is
identified as the boundary of the standard shadow, which shrinks in size with
increasing $Q$. For the HBHs with $Q=1.07$ and $Q=1.074$, the photon ring is
wide enough to leave imprints on the blurred images, where brighter and wider
blurred rings are observed.}%
\label{fig: blurredImage}%
\end{figure}

So far, we have considered high resolution images of HBHs surrounded by an
accretion disk, which present some interesting features of the photon ring. To
gain some insight into the effects of the photon ring on a realistic
observation, we blur the images of the accretion disk in Figs.
\ref{figure single peak 1}-\ref{figure double peak 2} with a Gaussian filter
with standard derivation equal to 1/12 the field of view to mimic the EHT
resolution \cite{Gralla:2019xty}, as shown in the lower row of Fig.
\ref{fig: blurredImage}. In Fig. \ref{fig: blurredImage}, the white dotted
circles represent critical circles, whose radii are the impact parameters of
photon spheres, and the standard shadow is defined as the region inside the
critical circle or that of smaller radius if there exist two photon spheres.

In the first column of Fig. \ref{fig: blurredImage}, we display the high
resolution and blurred images of the HBH from Fig. \ref{figure single peak 1},
which is in the single-peak I family. The blurred image is primarily
determined by the direct emission and the lensing ring, and has a blurred
bright ring at the location of the lensing ring, which indicates that the
observed shadow almost coincides with the standard shadow. In the second
column, we exhibit the high resolution and blurred images of the HBH of the
double-peak II family from Fig. \ref{figure double peak 1}, which has two
critical circles of similar radii. Compared with the single-peak I family, the
existence of two photon spheres slightly increases the brightness of the
blurred bright ring. In addition, the observed shadow in the blurred image is
almost same as the standard shadow.

On the other hand, when the photon ring is wide enough (e.g., Figs.
\ref{figure single peak 2} and \ref{figure double peak 2}), its effects can
become quite noticeable in the blurred images. Indeed, the high resolution and
blurred images of the HBH of the double-peak II family from Fig.
\ref{figure double peak 2} are shown in the third column of Fig.
\ref{fig: blurredImage}. In this case, the radii of the two critical circles
are quite different, which significantly increases the width of the photon
ring. Although the internal structure of the photon ring is washed out by the
blurring, the wide photon ring leads to a brighter and wider blurred ring than
in the single-peak I family. Since the inner edge of the blurred bright ring
is at the smaller critical circle, the observed shadow nearly matches the
standard shadow. In the last column, the high resolution and blurred images of
the HBH of the single-peak II family from Fig. \ref{figure single peak 1} are
presented. The ankle-like structure of the effective potential also notably
increases the brightness and width of the blurred bright ring. Moreover, the
blurred bright ring is at the location of the ankle-like structure, and hence
has a large radius than the critical circle, which means that the observed
shadow is larger than the standard shadow.

\section{Conclusion}

\label{sec:Discussion-and-conclusion}


In this paper, we investigated the behavior of null geodesics and
observational appearance of a HBH surrounded by an optically and geometrically
thin accretion disk in the EMS model with an exponential scalar coupling.
Depending on the profile of the effective potentials for photons, we found
that the HBH solutions are categorized into four different families, namely
single-peak I, single-peak II, double-peak I and double-peak II. The
single-peak I family dominates the allowed parameter space, while the other
three families exist in a small region with small $\alpha$ and large $Q$. For
the single-peak I solution, the photon ring is very narrow and makes little
contribution to the total brightness. Therefore, the accretion disk images
bear much similarity to these of various static black holes with a single-peak
potential \cite{Gralla:2019xty,Zeng:2020vsj,Peng:2020wun,He:2021htq}, e.g., a
bright ring due to the lensing ring shows up (Fig. \ref{figure single peak 1}%
). On the other hand, an additional ankle or peak-like structure emerges in
the single-peak II and double-peak II solutions, and results in the appearance
of a new finite or infinite peak in the $n(b)$ curve. Consequently, the photon
ring can become sufficiently broader, and its inner structure comes into play,
resulting in two bright annuluses in the observed image (Figs.
\ref{figure single peak 2} and \ref{figure double peak 2}). The annulus of
smaller radius is quite narrow, while the one of larger radius is much wider,
comprising multiple concentric thin bight rings with different luminosity. To
illustrate effects of the photon ring on observations, we presented the images
of the HBHs at low resolutions, roughly corresponding to the EHT resolution
(Fig. \ref{fig: blurredImage}). For the single-peak II and double-peak II
solutions, the details of bright annuluses are washed out, giving a
considerably bright and wide ring in the blurred image.

We end with some comments on future studies from several perspectives. In this
paper, we considered the concept \textquotedblleft photon
ring\textquotedblright, which is defined as the collection of all $m\geq3$
bands \cite{Gralla:2019xty}. Since we observed the complicate inner structure
inside the wide photon ring (Figs. \ref{figure single peak 2} and
\ref{figure double peak 2}), it would be interesting to study this rich
structure by performing detailed analysis on each $m\geq3$ band. Moreover, the
ankle- and peak-like structures in the effective potentials in the
aforementioned two cases behave like plateaus (upper-left panels of Figs
\ref{figure single peak 2} and \ref{figure double peak 2}), which is
reminiscent of the potential profiles reported in
\cite{Tsukamoto:2020iez,Tsukamoto:2020uay}. It suggests that a marginally
unstable photon sphere is likely to appear during the transition between the
single-peak II and double-peak II families. In addition, apart from the
exponential coupling of the EMS model considered in this paper, exploring
other coupling functions may provide us more novel phenomena about the photon
spheres and disk images. Finally, it would gain more insights into the
modified gravity by considering a rotating HBH surrounded by an accretion disk
and comparing it with black hole images released in future observations, such
as the Next Generation Very Large Array \cite{2015IAUGA..2255106H}, the Thirty
Meter Telescope \cite{Skidmore:2015lga}, and the BlackHoleCam
\cite{Goddi:2017pfy}.

\begin{acknowledgments}
We thank Guangzhou Guo and Li Li for his helpful discussions and suggestions.
This work is supported in part by NSFC (Grant No. 11875196, 11375121, 11947225
and 11005016).
\end{acknowledgments}

\appendix

\section{Light Propagation in the EMS Model}

In this appendix, we use the geometric optics approximation
\cite{Misner:1974qy,Fedderke:2019ajk,Schwarz:2020jjh} to derive the equations
of geometric optics governing light propagation in the EMS model with the
action
\begin{equation}
S=\int d^{4}x\sqrt{-g}\left[  \mathcal{R}-2\partial_{\mu}\phi\partial^{\mu
}\phi-e^{\alpha\phi^{2}}F_{\mu\nu}F^{\mu\nu}\right]  .
\end{equation}
The equation of motion for the electromagnetic field $A^{\mu}$ is then given
by
\begin{equation}
\partial_{\mu}\left[  \sqrt{-g}e^{\alpha\phi^{2}}F^{\mu\nu}\right]
=0.\label{eq:eomF}%
\end{equation}
Assuming the wavelength of the photon is much smaller than the other physical
scales of the system (i.e., the geometric-optics limit), we consider the
ansatz%
\begin{equation}
A^{\mu}=\operatorname{Re}\left[  \mathcal{A}^{\mu}\exp\left(  i\psi\right)
\right]  ,\label{eq:ansatz}%
\end{equation}
where $\mathcal{A}^{\mu}$ is the slowly evolving amplitude, and $\psi$ is a
rapidly oscillating function of time and space. The wave vector $k_{\mu}$ of
light rays is identified\ as%
\begin{equation}
k_{\mu}\equiv\partial_{\mu}\psi.
\end{equation}
Putting the ansatz $\left(  \ref{eq:ansatz}\right)  $ into Eq. $\left(
\ref{eq:eomF}\right)  $, we obtain the term of order $k^{2}$, which is the
leading term in the eikonal approximation,%
\begin{equation}
k^{\mu}k_{\mu}=0\text{,}%
\end{equation}
where the Lorentz gauge $\nabla_{\mu}A^{\mu}=0$ is used. Acting on the above
equation with a covariant derivative leads to%
\begin{equation}
k^{\mu}\nabla_{\mu}k_{\nu}=0,\label{eq:keqn}%
\end{equation}
where we use $\nabla_{\mu}k_{\nu}=\nabla_{\nu}k_{\mu}$. From Eq. $\left(
\ref{eq:keqn}\right)  $, it shows that light rays of the EMS model are
described by null geodesics in the geometric-optics limit.

\bibliographystyle{unsrturl}

\end{document}